\documentclass{eas}
\usepackage{graphicx}
\def\lsim{\lower.5ex\hbox{$\; \buildrel < \over \sim \;$}}
\def\gsim{\lower.5ex\hbox{$\; \buildrel > \over \sim \;$}}
\begin{document}

\title{X- and Gamma-Ray Line Emission Processes} 
\author{V. Tatischeff}\address{Centre de Spectrom\'etrie Nucl\'eaire et de
Spectrom\'etrie de Masse, IN2P3-CNRS, 91405 Orsay, France}
%
%
\begin{abstract}
This chapter is intended to provide a general presentation of the atomic and
nuclear processes responsible for X-ray line and gamma-ray line emission in
various astrophysical environments. I consider line production from hot
plasmas, from accelerated particle interactions, from the decay of
radioactive nuclei synthesized in stars and from positron annihilation.
Spectroscopic properties of these emissions are discussed in the light of
the detection capabilities of modern space instruments. 
\end{abstract}
\maketitle
\section{Introduction}

X- and gamma-ray emission lines are valuable signatures of various
high-energy processes at work in the universe, including heating of
astrophysical gas to very high temperatures, particle acceleration and 
nucleosynthesis. Perhaps the most important realization of pioneering 
high-energy astronomy was that hot plasmas are found essentially everywhere 
in the universe. More and more detailed observations of thermal X-ray lines 
provide detailed and often unique information on a wide variety of 
astrophysical sites, including stellar environments, supernova explosions, 
accreting compact objects, interstellar and intergalactic media and active 
galactic nuclei. 

Particle acceleration is believed to occur in most of these sites. 
Our understanding of acceleration mechanisms should greatly 
benefit from the detection of lines produced through atomic and nuclear 
interactions of energetic particles with ambient matter. The sun provides 
valuable examples of these nonthermal radiation processes. The study of 
nuclear de-excitation lines produced in solar flares has now become a proper 
domain of solar physics. 

Gamma-ray lines are also emitted in the decay of radioactive nuclei.
Measurements of gamma-ray activities from cosmic radionuclei testify to
the ongoing synthesis of chemical elements and their isotopes in the Galaxy
and beyond. These observations are now cornerstones for models of 
nucleosynthesis in novae, supernovae and stellar interiors. 

Positron-electron annihilation radiation is relevant to almost all of
high-energy astrophysics. Positrons can be produced by various pair-creation
processes in relativistic plasmas, by accelerated particle interactions and 
by the $\beta^+$-decay of radioisotopes. Observations of their annihilations
offer a unique vision of high-energy accreting sources,
nucleosynthesis sites and the global structure of the Milky Way.

The aim of this text is to present the physical processes responsible for 
these various line emissions. I also consider emission processes 
which have not been observed with assurance yet: X-ray line production from 
accelerated particle interactions and from the decay of cosmic 
radioactivities, and gamma-ray line emission from thermonuclear 
plasmas. However, given the advent of new X-ray and gamma-ray 
satellites with unprecedented sensitivities and spectral resolutions 
(see Barret and Kn\"odlseder, this volume), I am optimistic that at 
least some of these emissions could be detected in the near future. 

The scope of this review is restricted to emission lines. In particular, I do
not discuss X-ray absorption lines recently observed from a handful of active 
galactic nuclei, as well as cyclotron absorption features in intense magnetic
fields, detected from several X-ray binaries. Nor do I treat the 
high-energy gamma-ray emission from pion decay. 

The plan of this review is the following: in section \S~2, I present
the basic processes of X-ray line emission in both thermally ionized and
photoionized plasmas; in \S~3, I consider nonthermal X-ray line production 
in interactions of accelerated electrons and ions with ambient gas; in \S~4,
I deal with gamma-ray line emission from nuclear collisions and discuss 
thermonuclear reactions as well as accelerated ion interactions; in \S~5, I 
consider both the gamma-ray lines and the X-ray lines emitted by the decay
of radioactive nuclei synthesized in stars; finally, positron 
annihilation radiation is discussed in \S~6. 

\section{X-ray line emission from hot plasmas}

Thermal X-ray emission is observed from a great variety of sources in the
universe (see, e.g., Astronomy \& Astrophysics, volume 365, for an
overview of the recent {\it XMM-Newton} observations). It
is produced in various ion-electron plasmas, which can be first 
classified as a function of their optical depth in the X-ray energy range 
(0.1$\le$$E_X$$\le$100 keV):
\begin{equation}
\tau \equiv \tau(E_X)=\alpha(E_X) \times D~,
\end{equation}
where $\alpha(E_X)$ is the linear absorption coefficient (in c.g.s. units of
cm$^{-1}$) at photon energy $E_X$ and $D$ is the typical dimension of the 
emitting plasma (cm)\footnote{~$\alpha(E_X)$ is also called the
{\it extinction coefficient} when it includes Compton scattering of the X-rays
against the free electrons (e.g. Rybicki \& Lightman \cite{ryb79}).}. 
Optically thick sources ($\tau$$>>$1) such as isolated neutron stars 
essentially emit blackbody radiation, with some possible discrete spectral 
structures due to complicated radiation transfer effects (e.g. Rajagopal \& 
Romani \cite{raj96}). Optically thin sources include stellar coron\ae,
supernova remnants, superbubbles, the hot interstellar medium, clusters of
galaxies and the intergalactic medium. The emission properties of such
sources are usually discussed in terms of the {\it coronal model}, based on
the pioneering attempt of Elwert (\cite{elw52}) to explain the X-ray emission
of the solar corona. In this model, the plasma radiation is essentially
due to collisions of thermal electrons with ions, but the mechanism
for heating the electrons to temperature $T_e$ typically $>$~10$^6$~K is 
not specified, i.e. $T_e$ is an external parameter of the model.

Different X-ray emitting plasmas are found in astrophysical objects such as 
X-ray binaries and active galactic nuclei, in which a nebular gas of
relatively high density can be strongly illuminated by the compact X-ray 
source. The physical state and radiation of such photoionized plasmas are 
generally assumed to be primarily controled by the emission of the compact 
source and are usually described in terms of the {\it nebular model}, which 
owes its name to the resemblance of these objects with planetary nebulae. In 
this section, I briefly present the X-ray line production in the coronal model 
(\S~2.1) and in the nebular model (\S~2.2) and then discuss some diagnostics 
of plasma parameters based on high-resolution, X-ray line spectroscopy 
(\S~2.3).

\subsection{Thermally ionized plasmas: the coronal model}

The coronal model (see Mewe \cite{mew99} for a comprehensive description) 
applies to an optically thin, tenuous, thermal plasma, in which (i) the 
electrons and the ions are relaxed to Maxwellian energy distributions, (ii) 
the ions are essentially in their ground state, i.e. excited state
configurations can generally be neglected for the calculation of the
ionization balance and (iii) each X-ray photon emitted in a microscopic 
process does not further interact with the ambient ions and electrons. First 
order deviations from the coronal model are discussed in Raymond 
(\cite{ray88}). 

\subsubsection{Ionization structure}

The ionization structure of each element of chemical symbol $Z$ and atomic
(proton) number $z$ is generally obtained by solving a set of $z$+1 coupled 
equations:
\begin{equation}
{1 \over n_e}{d\eta_Z^i \over dt} = \eta_Z^{i-1} C_{Z^{+(i-1)}} + 
\eta_Z^{i+1} \alpha_{Z^{+(i+1)}} - \eta_Z^i (C_{Z^{+i}}+\alpha_{Z^{+i}})~,
\end{equation}
where $\eta_Z^i$ is the ionic fraction of ion $Z^{+i}$ 
($\sum_{i=0}^z \eta_Z^i$=1), $C_{Z^{+i}}$ and
$\alpha_{Z^{+i}}$ (in c.g.s. units of cm$^3$ s$^{-1}$) are the total 
ionization ($Z^{+i}$$\rightarrow$$Z^{+(i+1)}$) and total recombination
($Z^{+i}$$\rightarrow$$Z^{+(i-1)}$) rate coefficients, respectively, and 
$n_e$ is the electron density (cm$^{-3}$). Note that the above equation is 
valid only if multiple ionizations can be neglected. If the ionization 
balance is in a steady-state equilibrium ($d\eta_Z^i/dt$=0), the ionization 
structure of element $Z$ is simply given by a set of equations connecting 
pairs of adjacent ionization states:
\begin{equation}
\eta_Z^i C_{Z^{+i}} = \eta_Z^{i+1} \alpha_{Z^{+(i+1)}}~.
\end{equation}
The plasma is said to be at coronal equilibrium when all elements are 
at ionization equilibrium. 

Ionizations are induced by thermal electron collisions. The ionization
coefficient $C_{Z^{+i}}$ is obtained by averaging the collisional
ionization cross section $\sigma^I_{Z^{+i}}(v)$ over the Maxwell-Boltzmann
distribution $f(v)$ of the relative collision velocity, which is generally
very close to the velocity distribution of the much lighter electrons: 
\begin{equation}
C_{Z^{+i}} = <\sigma^I_{Z^{+i}}(v) v> = \int_0^\infty v \sigma^I_{Z^{+i}}(v) 
f(v) dv~,
\end{equation}
with
\begin{equation}
f(v) = \bigg({2 \over \pi}\bigg)^{1/2} \bigg({m_e \over kT_e}\bigg)^{3/2} 
v^2 \exp\bigg[-{m_ev^2 \over 2kT_e}\bigg]~,
\end{equation}
where $m_e$ and $T_e$ are the electron mass and temperature and $k$ is the
Boltzmann constant. The ionization cross sections are estimated from 
laboratory measurements combined with atomic physics theory (in particular 
distorted wave calculations, e.g. Younger \cite{you82} and references 
therein).

\begin{figure}
\center
\includegraphics[width=8.4cm]{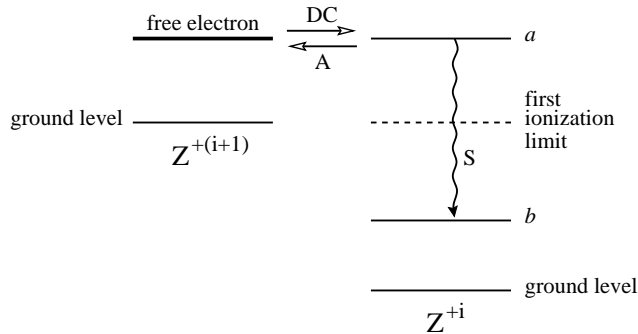}
\caption{Scheme of the dielectronic recombination process: a
dielectronic capture (DC) into a doubly-excited state {\it a}, followed by a 
stabilizing radiative transition (S) which can produce a satellite line (see
text). A: Autoionization.} 
\end{figure}

Apart from direct ionization ($Z^{+i}$+$e^-$$\rightarrow$$Z^{+(i+1)}$+2$e^-$),
ion $Z^{+i}$ can be stripped of one bound electron through the two-step 
process of excitation-autoionization
($Z^{+i}$+$e^-$$\rightarrow$$(Z^{+i})^*$+$e^-$$\rightarrow$$Z^{+(i+1)}$+2$e^-$),
which is due to the collisional excitation of an {\it inner-shell} electron 
to a bound level above the first ionization threshold, followed by
autoionization. The latter process is due to electron-electron
interactions, which can break the compound excited state $(Z^{+i})^*$ into its 
two components: ion $Z^{+(i+1)}$ and a free electron. The contribution of
excitation-autoionization to the total ionization cross section is 
calculated from the sum over all possible autoionizing levels {\it a} of the 
collisional excitation cross sections $\sigma^{E,a}_{Z^{+i}}$, weighted 
by the autoionization probability of each excited state. 

Recombination of a free electron can proceed either through a radiative
free-bound transition ($Z^{+(i+1)}$+$e^-$$\rightarrow$$Z^{+i}$+$h\nu$) or by 
a radiationless dielectronic recombination (DR). Charge-transfer
reactions are generally negligible for the hot plasmas considered in
the coronal model, but can be very important for photoionized plasmas
(\S~2.2). Since radiative recombination (RR) is the inverse of photoionization,
the rates of the two processes are connected through the principle of
time-reversal invariance. The application of this principle by means of
the detailed balance theorem leads to the Milne equation (Milne
\cite{mil24}), which is generally used to calculate RR 
rate coefficients from photoionization data (an example of the use of the
Milne equation is provided in \S~2.2.2). 

However, DR is often the dominant recombination process (e.g. Arnaud \& 
Rothenflug \cite{arn85}). It is schematically represented in Figure~1. The 
first step is a dielectronic capture (DC), in which the 
energy lost by the plasma electron is expended in the simultaneous excitation 
of one of the bound electrons in the core of the recombining ion $Z^{+(i+1)}$. 
The atomic level {\it a} thus produced is said to be doubly-excited, because 
its configuration involves two electrons occupying subshells above those which 
they would occupy in the ground state configuration of ion $Z^{+i}$. 
DC is the inverse of autoionization, so that 
the DC rates can be calculated from the autoionization rates, by applying the 
principle of detailed balancing. DR occurs if, 
instead of autoionizing, the doubly-excited state decays by a radiative 
transition to a level {\it b} below the first ionization limit (Fig.~1). The 
stabilizing decay of the core electron produces a so-called 
{\it satellite line}, as it is slightly shifted, usually to the 
long-wavelength side of the corresponding transition in ion $Z^{+(i+1)}$, due 
to the electrostatic shielding induced by the spectator outer electron. 
Satellite line emission can provide very powerful diagnostics of plasma 
parameters (\S~2.3). 

\begin{figure}
\center
\includegraphics[width=10.cm]{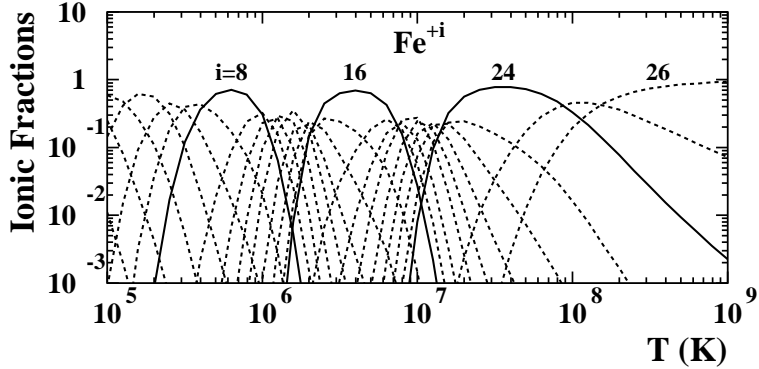}
\caption{Iron ionic fractions at ionization equilibrium as a function of 
plasma temperature (from Arnaud \& Raymond \cite{arn92}).}
\end{figure}

Shown in Figure~2 are the results of  Arnaud \& Raymond (\cite{arn92}) for 
the ionic fractions of Fe at ionization equilibrium. We see that ions with
closed-shell configurations are more stable than those with
partially filled L or M shells. Thus, He-like Fe~XXV (Fe$^{+24}$), Ne-like
Fe~XVII (Fe$^{+16}$) and Ar-like Fe~IX (Fe$^{+8}$), whose ground state
configurations are 1$s^2$, 2$s^2$2$p^6$ and 3$s^2$3$p^6$ respectively, are
dominant in large temperature ranges, because their ionization and DR rates 
are relatively low compared to those of adjacent ions. 

\subsubsection{Line emission}

In the coronal model, the line spectrum is dominated by radiative decays
following electron impact excitation, plus a smaller 
contribution of recombination lines. Collisional inner-shell ionization can 
however be important for plasma diagnostics based on fine spectroscopic 
analyses (\S~2.3). Considering only the dominant process of collisional 
excitation, the volume emissivity $P^{ab}_{Z^{+i}}$ (in units of photons 
cm$^{-3}$ s$^{-1}$) of a particular line transition $a \rightarrow b$ in ion 
$Z^{+i}$ can be written as
\begin{equation}
P^{ab}_{Z^{+i}} = n_e n_H a_Z \eta_Z^i S^{ga}_{Z^{+i}} B_{ab}~.
\end{equation}
Here, $n_H$ is the H atomic number density (cm$^{-3}$), $a_Z$ is the 
abundance of element $Z$ relative to H, $S^{ga}_{Z^{+i}}$ (cm$^3$ s$^{-1}$) is 
the rate coefficient for electron impact excitation of ion $Z^{+i}$ from its 
ground state to its excited state $a$ and $B_{ab}$ is the radiative branching 
ratio of the transition $a \rightarrow b$ among all possible transitions 
from level $a$:
\begin{equation}
B_{ab} = { A_{ab} \over \sum_{c<a} A_{ac}}~,
\end{equation}
where $A_{ab}$ is the transition probability (s$^{-1}$) of the spontaneous
decay $a \rightarrow b$ (collisional de-excitation is generally neglected). 
The excitation rate coefficients $S^{ga}_{Z^{+i}}$ are calculated by averaging 
the corresponding cross sections over the Maxwellian electron velocity 
distribution (see eq.~2.4). Semi-empirical fitting formulae have been derived 
for the most significant exciting transitions in all relevant ions 
(Mewe \cite{mew99} and references therein). 

The line spectrum is generally dominated by the ``allowed'' electric dipole
($E1$) transitions\footnote{~In the Russell-Saunders $LS$-coupling, which
generally holds for the most abundant ions (i.e. whose atomic number
$z$$<$29), the quantum selection rules of the $E1$ transitions are (e.g.
Sobelman \cite{sob79}):
$\Delta J$=$J_a$-$J_b$=0 or $\pm$1, with $J_a$+$J_b$$\ge$1;
$\Delta M$=$M_{J_a}$-$M_{J_b}$=0 or $\pm$1; $\Delta L$=$L_a$-$L_b$=0 or
$\pm$1, with $L_a$+$L_b$$\ge$1; $\Delta S$=$S_a$-$S_b$=0; $\pi_a$=-$\pi_b$;
and $\Delta \ell$=$\ell$-$\ell'$=$\pm$1. Here, $L_a$ and $S_a$ are the
orbital and spin angular momenta of the atomic energy level $a$, which
combine to the total angular momentum $J_a$ of magnetic quantum number
$M_{J_a}$; $\pi_a$ is the parity of level $a$; and $\ell$ and $\ell'$ are the
orbital angular momenta of the individual ``jumping'' electron in the atomic
states $a$ and $b$, respectively. Radiative transitions which violate one of
these selection rules are said to be ``forbidden'' and are usually weaker.}.
However, forbidden lines can be relatively bright if they are produced by
the decays of excited levels which cannot de-excite by an $E1$ transition. 
With the advent of new X-ray instruments with high sensitivity and
resolution capabilities, such as those aboard the {\it XMM-Newton} and {\it
Chandra} satellites, the correct modeling of relatively weak lines has become 
increasingly important. The three main coronal-plasma emission codes, 
SPEX (Kaastra {\em et al.} \cite{kaa96}), CHIANTI (Landi {\em et al.}
\cite{lan99}) and APEC (Smith {\em et al.} \cite{smi01}), now calculate 
tens of thousands of lines in the X-ray energy band.

\begin{figure}
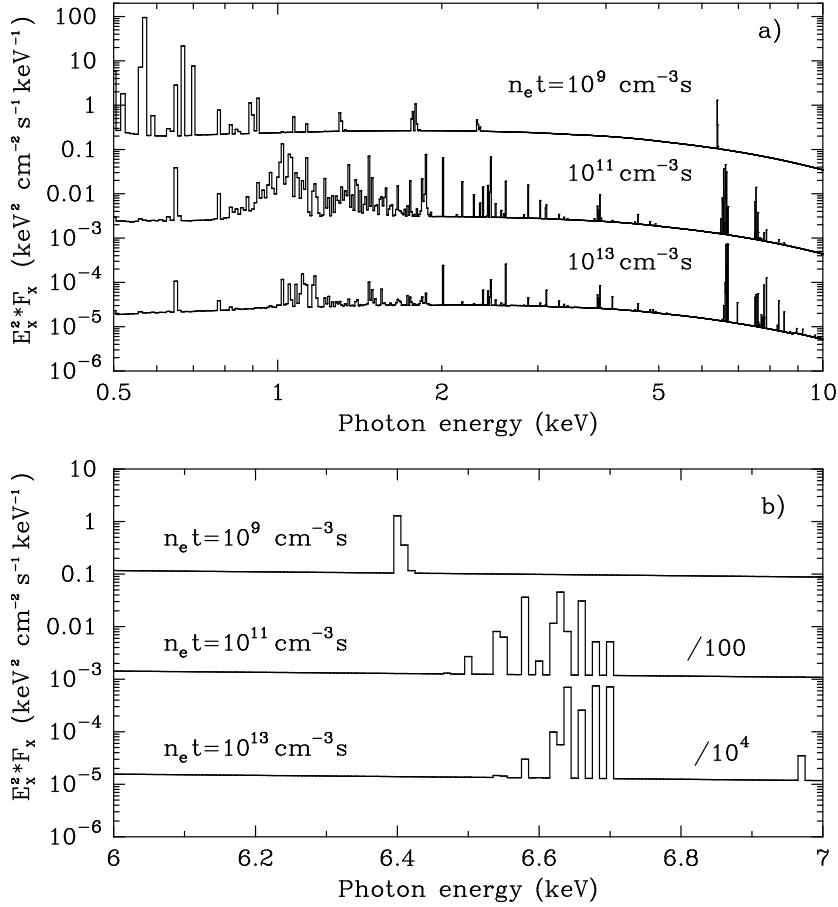

\begin{center}
\includegraphics*[width=11.cm]{thermx1p.eps}
\\
\includegraphics*[width=11.cm]{thermx2p.eps}
\end{center}
\caption{Thermal X-ray emissions between 0.5 and 10~keV (panel {\it a}) and
around the Fe K line complex (panel {\it b}) of an optically thin transient
plasma with solar abundances and electron temperature $T_e$=30~MK, for three
values of the ionization parameter. For $n_et$=10$^{13}$~cm$^{-3}$~s, the 
plasma has reached the ionization equilibrium. The calculations are normalized 
to the reduced emission measure $EM/(4\pi D^2)$=10$^{14}$~cm$^{-5}$, where 
$D$ is the distance to the source and $EM=\int n_e n_H dV$ (in cm$^{-3}$) is 
the total emission measure (which is directly proportional to the X-ray 
production rate, see eq.~2.6). In each panel, the spectra for $n_et$=10$^{11}$ 
and 10$^{13}$~cm$^{-3}$~s are divided by 100 and 10$^4$, respectively. All
spectra are plotted in energy bins of 10 eV.}
\end{figure}

As an illustration of the dependence of the X-ray line spectrum on the
ionization structure of a radiating plasma, Figure~3 shows calculated
emissions of a transient source, that has been instantaneously heated to
$T_e$=30$\times$10$^6$~K at time $t$=0. I used the non-equilibrium ionization
(NEI) plasma model (Borkowski {\em et al.} \cite{bor01} and references
therein), recently implemented in the XSPEC v11 spectral software package
(see Ballet, this volume). Assuming constant electron temperature and
density, we see from eq.~(2.2) that the evolution of the ionization
structure can be entirely characterized by the product $n_et$, which is
called the ionization parameter. The evolving line spectrum is superimposed
on a quasi-static continuum, which, for $T_e$=30~MK, is mainly produced by 
the thermal bremsstrahlung of the plasma electrons (free-free emission). 
However, the continuum radiation of plasmas with lower temperatures 
is often dominated by radiative recombinations (Mewe {\em et al.} 
\cite{mew86}). This emission is considered in \S~2.2.2.

The line emission generally shifts towards higher energies as the emitting 
ions become more and more ionized. For example, we see for 
$n_et$=10$^{11}$~cm$^{-3}$~s in panel ({\it a}), a prominent line forest at 
$\sim$1~keV, which mainly consists of Fe L lines, i.e. 
3$\ell$$\rightarrow$2$\ell'$ electronic transitions in ions Fe~XVII-XXIV. 
This emission is much weaker for $n_et$=10$^{13}$~cm$^{-3}$~s, because in a 
30~MK plasma at ionization equilibrium the most abundant Fe ion is the 
He-like Fe~XXV (see Fig.~2). The strongest line emission is then the Fe 
K$\alpha$ line complex (2$p$$\rightarrow$1$s$ transitions in the Siegbahn 
notation) at $\sim$6.7~keV. Fe K lines are also produced by inner-shell 
excitation or ionization of less ionized Fe, but as shown in panel ({\it b}), 
their energies are slightly lower, because the atomic levels get closer as 
the electrostatic shielding induced by the bound electrons increases. For 
example, the Fe K$\alpha$ line (which is made of the two very close components 
K$\alpha_1$$\equiv$2$p_{3/2}$$\rightarrow$1$s_{1/2}$ and 
K$\alpha_2$$\equiv$2$p_{1/2}$$\rightarrow$1$s_{1/2}$) is at
6.40~keV for Fe~II-XII, 6.70~keV for Fe~XXV and 6.97~keV for
Fe~XXVI (see panel {\it b}). Determinations of the precise energy and 
intensity of various lines thus provide direct information on the ionization 
structure of a coronal plasma.

\subsection{Photoionized plasmas: the nebular model}

The nebular model (e.g. Kallman \& McCray \cite{kal82}, Liedahl \cite{lie99}) 
is intended to describe the physical state and emission of a plasma 
illuminated by intense X-ray radiation. The most important applications of 
this model are accretion-powered X-ray sources: active galactic nuclei, X-ray 
binaries and cataclysmic variables. In these objects, circumsource material 
is believed to intercept some of the radiation emitted from a compact region 
near the accreting source and to reprocess it into line and continuum emission 
in other portions of the spectrum. The main assumption of the nebular model is 
that the temperature and ionization structure of the illuminated gas 
are essentially governed by the interactions of the radiation field with the 
ion-electron plasma. Possible additional heating sources, such as shocks or 
magnetic-field reconnection, are thus generally neglected. 

\subsubsection{Ionization structure and thermal equilibrium}

The temperature and ionization structure of a photoionized plasma are
obviously intimately connected. For example, recombination of thermal 
electrons affects both the charge state distribution and the plasma cooling 
rate. Thus, the state of the plasma has to be determined by solving a set 
of coupled equations describing the ionization balance together with the 
thermal equilibrium. Most photoionized-plasma models which have been 
developed so far (see Ferland {\em et al.} \cite{fer95} and references 
therein) assume that the ionization and energy balances are in a 
steady-state, i.e. that the time scales for variation in the compact source 
emission and in the plasma global parameters are long compared to the 
characteristic time scales for the atomic processes. The ionic fractions of 
adjacent charge states are 
then connected through a generalization of eq.~(2.3), which obviously has to 
take into account the dominant process of photoionization, but also the 
possible contribution of charge-exchange reactions (e.g. Halpern \& Grindlay 
\cite{hal80}):
\begin{equation}
\eta_Z^i \cdot [\zeta_{Z^{+i}}(F_\epsilon) + n_eC_{Z^{+i}}(T) + K^I_{Z^{+i}}(T)] 
= \eta_Z^{i+1} \cdot [n_e\alpha_{Z^{+(i+1)}}(T) + K^R_{Z^{+(i+1)}}(T)]~.
\end{equation}
Here, $\zeta_{Z^{+i}}(F_\epsilon)$ is the photoionization rate (in units of
s$^{-1}$), which depends on the flux of the illuminating radiation
$F_\epsilon$ (see below), and $K^I_{Z^{+i}}(T)$ and $K^R_{Z^{+i}}(T)$ are the
effective ionization and recombination rates due to charge-transfer
reactions (also in s$^{-1}$), which, for a gas of given composition and
density, only depend on the ionic fractions of H and He, and on the plasma
temperature $T$. Ionization by thermal electron impact (of rate coefficient
$C_{Z^{+i}}(T)$, \S~2.1.1) is often unimportant, because, contrary to coronal
plasmas, the temperature of photoionized gas may not reach the ionization
threshold energy of many ions. Lower temperatures can also modify the
relative contributions of DR and RR to the total recombination rate
coefficient $\alpha_{Z^{+(i+1)}}(T)$ (\S~2.1.1): in a photoionized
plasma, radiationless dielectronic capture proceeds primarily through core
excitations of relatively low energies, i.e. essentially transitions of
core electrons to the same shell ($\Delta n$=0 excitations, $n$ being the
principal quantum number). Thus, RR generally dominates the total 
recombination rate for H-like and He-like ions, since these species have no 
available $\Delta n$=0 excitations. On the other hand, DR can be the main 
recombination process for ions with partially filled L or M shells (Liedahl 
\cite{lie99} and references therein).  

The photoionization rate $\zeta_{Z^{+i}}(F_\epsilon)$ is obtained by 
folding the radiation field with the corresponding photoionization cross 
section $\sigma^{pI}_{Z^{+i}}(\epsilon)$:
\begin{equation}
\zeta_{Z^{+i}}(F_\epsilon) = \int_{I_{Z^{+i}}}^\infty {F_\epsilon(\epsilon,R) 
\over \epsilon} \sigma^{pI}_{Z^{+i}}(\epsilon) d\epsilon~,
\end{equation}
where $I_{Z^{+i}}$ is the ionization threshold energy of ion $Z^{+i}$ and 
$F_\epsilon(\epsilon,R)$ is the net energy flux at the distance $R$ from the 
compact source (in c.g.s units of erg s$^{-1}$ cm$^{-2}$ erg$^{-1}$). For 
example, for an isotropic point source of total luminosity $L_\epsilon$ and 
spectrum $dL_\epsilon/d\epsilon$=$L_\epsilon$$s(\epsilon)$, 
assuming that the source radiation is not significantly attenuated by 
photoabsorption, the photoionization rate is simply given by 
\begin{equation}
\zeta_{Z^{+i}}(F_\epsilon) = {L_\epsilon \over 4 \pi R^2} 
\int_{I_{Z^{+i}}}^\infty {s(\epsilon) \over \epsilon} 
\sigma^{pI}_{Z^{+i}}(\epsilon) d\epsilon~.
\end{equation}

However, photoionization is often enhanced by the emission of one or more 
Auger electrons in addition to the original photoelectron. For example,
photoionization of a K-shell electron in an ion having two or more L-shell
electrons leaves the resulting ion in an {\it autoionizing} state that can 
decay by ejecting at least one L-shell electron. The L-shell vacancy thus 
produced can further be filled by an other radiationless Auger transition if 
the ion has two or more M-shell electrons. This cascade process can lead to 
the ejection of up to eight Auger electrons following the creation of a 
K-shell vacancy in neutral iron (Kaastra \& Mewe \cite{kaa93}). Thus, 
ionization balance calculations in the framework of the nebular model are 
generally complicated by the possible coupling of several charge states in 
each element (Weisheit \cite{wei74}). Note that additional ionization can be
produced by collisions of the fast ejected electrons with neighbouring
ions (Halpern \& Grindlay \cite{hal80}). However, if most of ambient H and He 
are ionized, these electrons will slow down primarily by elastic scattering 
with thermal electrons, thus depositing their energy as heat. 

Charge-transfer reactions can be very important for the ionization
structure of a photoionized plasma, if a significant abundance of neutral H
or He atoms (i.e. with neutral fraction typically exceeding $\sim$0.01) 
coexists with ionized heavy elements. They are of the form 
\begin{equation}
Z^{+i} + A^0 \rightleftharpoons Z^{+(i-1)} + A^+ + \Delta E~,
\end{equation}
where $A$$\equiv$H or He and $\Delta E$ is the energy defect, which can be
either a positive or a negative quantity depending on the reaction and on
the atomic states involved in the electron transfer (e.g. Kingdon \& Ferland 
\cite{kin99}). The rate coefficients are obtained by integrating the reaction 
cross sections over the Maxwell-Boltzmann distribution of the relative
collision velocity. Detailed quantum mechanical calculations were performed 
for few important reactions. For other systems, the cross sections can be
reasonably estimated by use of the Landau-Zener approximation 
(Kingdon \& Ferland \cite{kin96} and references therein).

In the nebular model, the plasma temperature is determined by solving the
equation of thermal equilibrium:
\begin{eqnarray}
n_e \Gamma_e + n_H \sum_Z a_Z \sum_i \eta_Z^i (\Gamma^{pI}_{Z^{+i}} +
\Gamma^{CT}_{Z^{+i}}) =~~~~ \nonumber \\
n_e \Lambda_{ff} +  n_H \sum_Z a_Z \sum_i \eta_Z^i 
(\Lambda^{R}_{Z^{+i}} + \Lambda^{C}_{Z^{+i}} + \Lambda^{CT}_{Z^{+i}})~,
\end{eqnarray}
where the left- and the right-hand sides represent the total heating and
cooling rates (erg cm$^{-3}$ s$^{-1}$) of the thermal
bath, respectively. Here, $\Gamma_e$ is the heating rate per free electron
due to Compton scattering\footnote{~Compton scattering is generally the
dominant heating process of X-ray-emitting photoionized plasmas. The net rate
coefficient expressing the excess of Compton heating over Compton cooling can 
be written as (Ross \cite{ros79}):
\begin{equation}
\Gamma_e={\sigma_T \over m_e c^2} \bigg[\int_0^\infty \big(\epsilon - 
{21 \over 5}{\epsilon^2 \over m_ec^2}\big)F_\epsilon(\epsilon,R)d\epsilon - 
4kT\int_0^\infty F_\epsilon(\epsilon,R)d\epsilon\bigg]~,
\end{equation}
where $\sigma_T$ is the Thomson cross section and $c$ is the speed of 
light. The first term in the brackets represents the heating of relatively 
cold electrons by scattering with X-rays and includes the first order 
Klein-Nishina correction, and the second term represents the cooling of hot 
electrons by scattering with lower energy photons (see also Marcowith, this 
volume).}; $\Gamma^{pI}_{Z^{+i}}$ is the effective photoionization
heating rate, which includes heating by the Auger electrons (Shull 
\cite{shu79});
$\Gamma^{CT}_{Z^{+i}}$=$K^{R~{\rm or}~I}_{Z^{+i}}$$\times$$\Delta E$
is the heating rate due to exothermic ($\Delta E$$>$0) charge-transfer
recombinations or ionizations (Kingdon \& Ferland \cite{kin99}); $\Lambda_{ff}$
is the thermal bremsstrahlung cooling rate due to free-free emission (see 
Marcowith, this volume); $\Lambda^{R}_{Z^{+i}}$ is the recombination cooling 
rate due to both RR and DR (Halpern \& Grindlay \cite{hal80} and references 
therein); $\Lambda^{C}_{Z^{+i}}$ is the cooling rate due to collisional 
excitation or ionization by thermal electron impacts; and 
$\Lambda^{CT}_{Z^{+i}}$ is the cooling rate due to endothermic 
($\Delta E$$<$0) charge-transfer reactions. Alternatively, the thermal 
balance can be established by calculating the rate of removal or addition of 
energy to the local radiation field associated with each of the processes 
affecting atomic level populations (Kallman \& Bautista \cite{kal01}). 

\subsubsection{Line emission}

It is instructive to compare the line production in photoionized
plasmas and in thermally ionized plasmas. In the coronal model, the
electron population responsible for the plasma ionization structure is also
an efficient source of collisional line excitation. On the other hand, a
photoionized plasma is generally overionized relative to the electron
temperature, since the thermal electrons are not the dominant source
of ionization. Collisional excitation is then often unimportant in the
X-ray energy range and the line emission is preferentially due to 
recombination of relatively low-energy electrons and to photon-induced 
fluorescence. Therefore, whereas in the coronal model, line production in ion 
$Z^{+i}$ is generally proportional to its ionic fraction $\eta_Z^i$ 
(eq.~2.6), in the nebular model it is dominated by processes involving 
adjacent charge states: recombination of ion $Z^{+(i+1)}$ and inner-shell 
photoionization of ion $Z^{+(i-1)}$. This difference allows to distinguish 
between coronal and nebular gas emission of cosmic origin by observing 
various line intensity ratios from the same element. For example, Liedahl 
{\em et al.} (\cite{lie90}) have shown that the 
(3$s$$\rightarrow$2$p$)/(3$d$$\rightarrow$2$p$) line ratios from Fe XVII-XIX
can be used to discriminate between thermally ionized and photoionized
plasmas, because the 3$d$ lines are preferentially produced by electron
impact excitation in a hot coronal plasma (5-10~MK, see Fig.~2), whereas the 
3$s$ lines are mostly formed by recombination in a cooler ($\sim$0.1~MK) 
nebular plasma. 

Line emission in the X-ray energy range is dominated by the K lines of C, N,
O, Ne, Mg, Si, Ar, Ca, Fe and Ni and the L lines of Fe and Ni. Thus,
recombination lines are produced in relatively highly ionized ions, having
in their ground state at least one K-shell vacancy, or one L-shell vacancy
for Fe$^{+i}$ and Ni$^{+i}$. On the other hand, fluorescent lines being due 
to radiative transitions from the L or M shell following inner-shell 
photoionization, they are generally associated with colder, less ionized 
plasmas. 

Recombination radiation in overionized plasmas has been observed from 
several high mass X-ray binaries (e.g. Sako {\em et al.} \cite{sak99} and 
references therein). It consists of radiative recombination ``continua''
(see below) and lines produced through recombination to excited states 
followed by radiative cascades. The volume emissivity (photons cm$^{-3}$ 
s$^{-1}$) of a given line transition $a \rightarrow b$ in ion $Z^{+i}$ can be 
written as 
\begin{equation}
P^{ab}_{Z^{+i}} = n_e n_H a_Z \eta_Z^{i+1} \alpha^{a}_{Z^{+(i+1)}} B_{ab}~,
\end{equation}
where $\alpha^{a}_{Z^{+(i+1)}}$ is the effective recombination rate
coefficient to the level $a$ (including recombination to higher levels
followed by radiative transitions to level $a$) and the other quantities are
defined above in eq.~(2.6). Radiative recombination continuum (RRC) is produced
by the first free-bound transition. The monochromatic specific emissivity
(photons s$^{-1}$ cm$^{-3}$ erg$^{-1}$) from radiative capture of a
free electron of velocity $v$ is given by 
\begin{equation}
{dP^{RRC}_{Z^{+i}} \over d\epsilon}(\epsilon) = n_e n_H a_Z \eta_Z^{i+1} v f(v) 
\sigma^{RR}_{Z^{+(i+1)}}(v) {dv \over d\epsilon}~,
\end{equation}
where the energy of the radiated photon is equal to the sum of the
ionization potential of the recombined ion $(Z^{+i})^*$, which may or may not 
be formed in an excited state, plus the kinetic energy of the recombining 
electron, 
\begin{equation}
\epsilon = I_{(Z^{+i})^*} + {1 \over 2} m_e v^2~,
\end{equation}
$f(v)$ is the thermal electron velocity distribution (eq.~2.5) and 
$\sigma^{RR}_{Z^{+(i+1)}}(v)$ is the radiative recombination cross section,
which is related to the corresponding photoionization cross section through
the Milne equation (see e.g. Mewe \cite{mew99}):
\begin{equation}
\sigma^{RR}_{Z^{+(i+1)}}(v) = {g_{(Z^{+i})^*} \over g_{Z^{+(i+1)}}} \cdot
{\epsilon^2 \over (m_evc)^2} \cdot \sigma^{pI}_{(Z^{+i})^*}(\epsilon)~.
\end{equation}
Here, $g_{Z^{+i}}$=2$J_{Z^{+i}}$+1 is the statistical weight of ion
$Z^{+i}$, whose total angular momentum is $J_{Z^{+i}}$. Combining these
equations provides the RRC emissivity as a function of the electron
temperature:
\begin{equation}
{dP^{RRC}_{Z^{+i}} \over d\epsilon}(\epsilon) = \sqrt{{2 \over \pi}} \cdot
{g_{(Z^{+i})^*} \over g_{Z^{+(i+1)}}} \cdot {n_e n_H a_Z
\eta_Z^{i+1} c \sigma^{pI}_{(Z^{+i})^*}(\epsilon)
\epsilon^2 \over (m_e c^2 kT_e)^{3/2}} \cdot
\exp\bigg[-{(\epsilon-I_{(Z^{+i})^*}) \over kT_e}\bigg]~.
\end{equation}
The photoionization cross sections are rapidly decreasing functions of photon
energy from ionization thresholds. For example, they are about proportional to
$\epsilon^{-3}$ for hydrogenic ions (see also Fig.~6a). We see from the
above equation that the RRC emissivities are steep functions of energy 
as well. In photoionized plasmas, the RRC emissions often appear as 
relatively narrow features in the spectrum (sometimes ``line-like'' features), 
with approximate width $\Delta \epsilon$$\sim$$kT_e$ (Liedahl \cite{lie99}). 
Analyses of RRC shapes can thus allow determination of the electron 
temperature of the emitting plasma.  

Fluorescent line emission has been observed from a number of Galactic X-ray
binaries (e.g. Ebisawa {\em et al.} \cite{ebi96}, Sako {\em et al.} 
\cite{sak99}) and active galactic nuclei (Fabian {\em et al.} \cite{fab00}
and references therein). It is produced when an inner-shell photoionization is
followed by a radiative transition. The volume emissivity of a fluorescent
line $a \rightarrow b$ is thus given by 
\begin{equation}
P^{ab}_{Z^{+(i+j)}} = n_H a_Z \eta_Z^{i-1} \zeta_{Z^{+(i-1)}}^{a'}(F_\epsilon) 
\omega_{Z^{+i}}^{ab}~, 
\end{equation}
where $\zeta_{Z^{+(i-1)}}^{a'}(F_\epsilon)$ is the inner-shell photoionization
rate of ion $Z^{+(i-1)}$, leading to ion $Z^{+i}$ in the excited level $a'$
(see eqs.~2.9 and 2.10) and $\omega_{Z^{+i}}^{ab}$ is the $a \rightarrow b$
line fluorescence yield, i.e. the number of photons emitted in the line
$a \rightarrow b$ per ion $Z^{+i}$ in the level $a'$. Note that the line
$a \rightarrow b$ is not necessarily produced in ion $Z^{+i}$, since the
first decay of the level $a'$ can be an Auger transition to an excited state
$a''$ of ion $Z^{+(i+1)}$. Auger electron emission can repeat several times
before the line $a \rightarrow b$ is emitted in ion $Z^{+(i+j)}$ and the
calculation of the overall process can become cumbersome, since millions of
transitions can follow a single inner-shell ionization. For example, 5.1
millions of Auger or radiative transitions are possible after an Fe I
K-shell ionization (Kaastra \& Mewe \cite{kaa93}). The convention
is therefore to refer to the inner-shell vacancy produced in ion
$Z^{+(i-1)}$, rather than to the excited atomic level $a'$ of ion $Z^{+i}$.
For example, we speak of the K$\alpha$ fluorescence yield for {\it neutral}
Fe (for which I will adopt the notation
$\omega_{Fe}^{K\alpha}$$\equiv$$\omega_{Fe^{+0}}^{K\alpha}$), although 
this quantity strictly describes an atomic property of ion Fe$^{+1}$. 

The fluorescence yield is a strongly increasing function of ion atomic number
$z$. For example, $\omega_Z^K$=$\omega_Z^{K\alpha}$+$\omega_Z^{K\beta}$
varies approximately as $z^{3.25}$ for a K-shell vacancy produced in neutral
atoms up to Fe (Krause \cite{kra79}). That explains why the most prominent
fluorescent line observed so far is the K$\alpha$ line from weakly-ionized
Fe, since this element has the highest product of K$\alpha$ fluorescence yield
($\omega_{Fe^{+i}}^{K\alpha}$=0.305$\pm$0.001 for Fe I-IX, Kaastra \& Mewe 
\cite{kaa93}) and cosmic (i.e. solar) abundance 
($a_{Fe}$=3.2$\times$10$^{-5}$, Anders \& Grevesse \cite{and89}). The line 
emission at $\sim$6.4~keV is often accompanied by a sharp absorption edge at 
7.1 keV, the K-shell electron binding energy in neutral or weakly-ionized Fe: 
nearly 30\% (=$\omega_{Fe^{+i}}^{K\alpha}$) of the photons absorbed above 
this threshold energy are ``re-emitted'' in the Fe K$\alpha$ line. 

However, if the nebular plasma is optically thick to the ionizing continuum
in some portions of the spectrum, radiative transfer is generally important 
and has to be taken into account from a geometrical model of the emitting
source (e.g. Fabian {\em et al.} \cite{fab00}). The problem of continuum and
line transfer at intermediate optical depth is a difficult one and is
generally tackled from approximate solutions to the full transfer equation
(e.g. Kallman \& Bautista \cite{kal01}). It is particularly relevant for
resonance lines, which correspond to allowed transitions to ground ionic 
levels and can be emitted following radiative recombination or collisional
excitation. Indeed, photons emitted in such lines can have a high
probability to be absorbed by the same transition which emitted them and
``re-emitted'' at a new frequency. This frequency redistribution can repeat
many times until the photon either escapes the gas or is destroyed by
photoabsorption (e.g. Faurobert \cite{fau86}). 

\begin{figure}
\begin{center}
\includegraphics[width=10.cm]{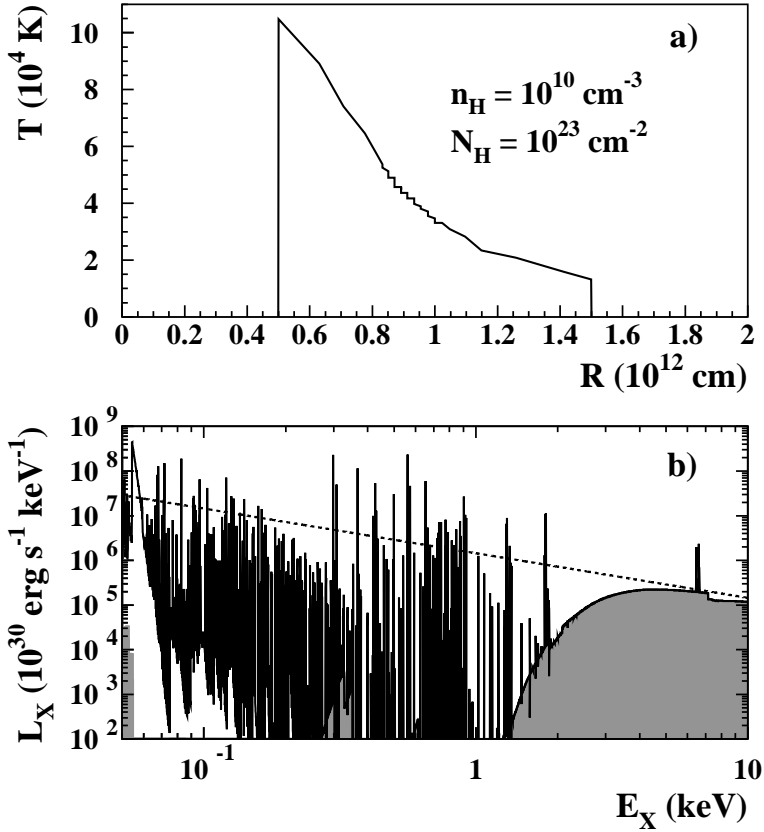}
\end{center}
\caption{Calculated thermal structure ({\it a}) and emission ({\it b}) of a 
spherical gas shell of solar composition and constant H density (irrespective 
of its charge state), $n_H$=10$^{10}$~cm$^{-3}$, illuminated by a steady 
state, isotropic point source at $R$=0. In panel ({\it b}), the dashed line 
is the source spectrum,
$L_X^{in}$$(E_X)$=1.45$\cdot$10$^{36}$$\times$$E_X$$^{-1}$~erg~s$^{-1}$~keV$^{-1}$, 
and the grey area shows the source emission after transmittance through the
shell. The total emerging spectrum (solid line) is the sum of the 
transmitted source emission plus the thermal radiation (mostly recombination 
radiation) and fluorescence emission from the photoionized shell.}
\end{figure}

By way of illustrated summary, I show in Figure~4 calculated thermal
structure and emission of a spherical gas shell photoionized by a central
source. I used the XSTAR version 2.1h computer code developed by Kallman
{\em et al.}\footnote{~See
http://heasarc.gsfc.nasa.gov/docs/software/xstar/xstar.html~.}. We see in
panel ({\it a}) that the temperature is found to decrease from $\sim$10$^5$~K
($kT$$\sim$8.6~eV) in the internal region of the shell to nearly 10$^4$~K in
the outermost region. Thus, collisional excitation is negligible in the
X-ray energy range and the line emission in panel ({\it b}) is mostly due to
radiative recombination. The Fe K$\alpha$ fluorescent line emission at
$\sim$6.5~keV and the corresponding absorption above 7.1~keV are also
visible. The irregular continuum emission from the photoionized shell is
produced by overlapping radiative recombination continua and two-photon
decays of metastable levels, which cannot decay by a single-photon
transition\footnote{~The simultaneous emission of two photons is important
for H- and He-like ions in the excited states 
$n$$^{2S+1}L_J$~$\equiv$~2$^2S_{1/2}$ and 2$^1S_0$, respectively. In 
this usual notation, $n$ is the principal quantum number of the outermost
occupied shell; $L$ is the total orbital quantum number, which, in the
Russell-Saunders $LS$-coupling, is given by the vectorial sum of the
individual electron orbital angular momenta; $S$ is the total spin angular
momentum, obtained by the vectoral coupling of the spin angular momentum
($s$=$1/2$) of all the electrons ($r$=$2S+1$ is called the multiplicity of
the level); and $J$ is the total angular momentum, which is the vector
sum $L$+$S$ for $LS$ coupling. The orbital quantum number $L$ is usually
designated by the symbols $S$, $P$, $D$, $F$, $G$\ldots (corresponding to
``\b{s}harp'', ``\b{p}rincipal'', ``\b{d}iffuse'', ``\b{f}undamental'' and
the subsequent letters to the alphabetical order) for $L$=0,1,2,3,4\ldots
Thus, the level 2$^2S_{1/2}$ of H-like ions corresponds to the electron 
being in the 2$s$ subshell and the level 2$^1S_0$ of He-like ions to the 
two electrons being in the 1$s$ and 2$s$ subshells with antiparallel spins 
($S$=0). It follows from the quantum selection rules that for these two 
levels, the 2$s$ electron cannot decay by a single-photon transition (e.g. 
Sobelman \cite{sob79}).} (thermal bremsstrahlung is negligible in this energy 
domain). The most prominent RRC in panel ({\it b}) is the triangular feature 
just above 54~eV  (i.e. in the extreme ultraviolet waveband), produced by 
free electron recombination to the K-shell of fully-ionized He.

\subsection{Plasma diagnostics from X-ray line emission}

As already mentioned, simple spectroscopic analyses of the X-ray line 
emission from a hot plasma can provide valuable information, such as the
ionization processes (photoionization and/or collisional ionization), the 
elemental abundances and the ionization balance. 
The recent launch of X-ray instruments with high sensitivity 
and spectral resolution, such as the grating spectrometers aboard the 
{\it XMM-Newton} and {\it Chandra} satellites, allows more refined
diagnostics of plasma parameters, among which those based on satellite lines 
and helium-like line triplets may deserve special attention. 

Satellite lines are produced by stabilizing transitions in the process of
dielectronic recombination (\S~2.1.1). The most important of such lines
arise from the dielectronic capture of a free electron into the L-shell and
appear on the long-wavelength side of the parent resonance line. For
example, the relatively strong satellite line
$j$$\equiv$1$s$2$p^2$[$^2D_{5/2}$]$\rightarrow$1$s^2$2$p$[$^2P_{3/2}$] is at
6.744~\AA~(1.838~keV) for Li-like Si~XII, whereas the corresponding resonance
line $w$$\equiv$1$s$2$p$[$^1P_1$]$\rightarrow$1$s^2$[$^1S_0$] in He-like
Si~XIII is at 6.647~\AA~(1.865~keV; Gabriel \cite{gab72}). The DR
satellite-to-resonance line intensity ratios are very sensitive to the
electron temperature of a coronal plasma. This is fundamentally due to the
resonant character of the DR process (Fig.~1) compared to the collisional 
excitation process: a DR satellite line is produced only by free electrons of 
energy close to the excitation energy of the satellite line level (within the 
small level width), whereas a resonance line can be excited by all electrons
with any energy above the resonance line energy. This also provides a way to
evaluate a possible deviation of the electron velocity distribution from a
Maxwellian distribution. Thus, the existence of a high-energy tail of 
nonthermal electrons in turbulent plasmas could be identified through an 
increase of the resonance line intensities relative to the DR satellite 
line intensities (Gabriel \& Phillips \cite{gab79}). 

Collisional inner-shell excitation (IE) can also contribute to the emission
of certain satellite lines, such as the line
$q$$\equiv$1$s$2$s$2$p$[$^2P_{3/2}$]$\rightarrow$1$s^2$2$s$[$^2S_{1/2}$] (at
6.718~\AA~in Si~XII). Because the simultaneous excitation of two
electrons is generally highly improbable, some lines ($j$ for example)
can only be produced from excited state configurations (1$s^2$2$p$ for line
$j$), which are negligible in low-density plasmas. However, these excited
levels can obtain a significant population at equilibrium for high 
electron density (e.g. at $n_e$$\sim$10$^{17}$cm$^{-3}$ for Si~XII), thus 
introducing a density dependence of satellite lines. 

The intensity ratio of a IE satellite line to its parent resonance line
is almost independent of temperature, since these two lines are excited by 
nearly the same electron population. However, it does depend on the plasma 
ionization structure, as the intensity of an IE satellite line in ion
$Z^{+i}$ is proportional to the ionic fraction $\eta_Z^i$, whereas the 
intensity of the corresponding resonance line in ion $Z^{+(i+1)}$ is 
proportional to $\eta_Z^{i+1}$. Thus, satellite lines can also provide 
diagnostics of the ionization state of a transient plasma. 

\begin{figure}
\begin{center}
\includegraphics[width=7.5cm]{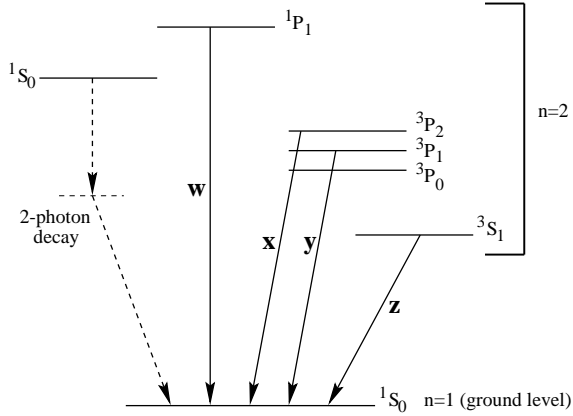}
\end{center}
\caption{Scheme of the first energy levels of He-like ions (simplified
Gotrian diagram). The He-like triplet is formed by the resonance line
{\it w}, the intercombination line {\it x}+{\it y} and the forbidden line
{\it z}. Also shown is the continuum-producing, two-photon decay of the
level 2$^1S_0$.}
\end{figure}

I now briefly present diagnostics based on the well-known He-like
$n$=2$\rightarrow$1 line triplet. This system consists of three close lines
(Fig.~5): the resonance $E1$ line $w$, the forbidden magnetic dipole ($M1$) 
line $z$ and the intercombination line $x$+$y$, which is made of 
two unresolvable components that violate the selection rule $\Delta S$=0 
(see footnote~2) because of magnetic interactions. In Si XIII for example, 
the lines are at 6.684~\AA~, 6.687~\AA~and 6.739~\AA, for the $x$, $y$ and 
$z$ transitions, respectively. As first shown by Gabriel \& Jordan
(\cite{gab69}), the emissivity ratio of the forbidden to intercombination
lines, 
\begin{equation}
R_{Z^{+i}} = {P^{z}_{Z^{+i}} \over P^{x}_{Z^{+i}}+P^{y}_{Z^{+i}}}~,
\end{equation}
is strongly sensitive to the electron density above a certain value
($\sim$5$\cdot$10$^{12}$ cm$^{-3}$ for Si~XIII and $\sim$5$\cdot$10$^{7}$ 
cm$^{-3}$ for C~V). This is due to the collisional excitation of the 
metastable level 2$^3S_1$ to the close levels 2$^3P_{0,1,2}$, which leads to 
a decrease of the $R_{Z^{+i}}$ ratio with increasing electron density. This 
density measurement can be applied to recombination dominated plasmas as well 
as to coronal plasmas. However, photoexcitation of the 2$^3S_1$ level to the
2$^3P$ term may not be negligible for photoionized plasmas illuminated by 
intense ultraviolet radiation (Porquet {\em et al.} \cite{por01b} and 
references therein). 

Gabriel \& Jordan (\cite{gab69}) also showed that the emissivity ratio
\begin{equation}
G_{Z^{+i}} = {P^{z}_{Z^{+i}} + (P^{x}_{Z^{+i}}+P^{y}_{Z^{+i}}) \over
P^{w}_{Z^{+i}}}
\end{equation}
can be used to diagnose the electron temperature of a coronal plasma, since
the collisional excitation rates have not the same temperature dependence
for the resonance line as for the forbidden and intercombination lines. This
ratio can also serve as an indication of the validity of the coronal model
(e.g. Porquet \& Dubau \cite{por00}). Indeed, a strong resonance line is
expected in plasmas dominated by collisional excitation. On the other hand,
since radiative recombinations to the 2$^3S$ and 2$^3P$ triplet terms are
favored as compared with the 2$^3P_1$ level, a significant contribution of
photoionization leads to a relatively high $G_{Z^{+i}}$ ratio (see also
Liedahl \cite{lie99}). Porquet \& Dubau (\cite{por00}) and Porquet {\em et
al.} (\cite{por01b}) have recently provided detailed calculations of the line
ratios $R_{Z^{+i}}$ and $G_{Z^{+i}}$ for the most important He-like ions. 

Satellite lines and He-like triplet lines have proven to be very useful to 
characterize laboratory plasmas, as well as coronal plasmas heated by solar 
flares (e.g. Harra-Murnion {\em et al.} \cite{har96}). The new generation of 
X-ray spectrometers now allows to use the same powerful diagnostics for 
cosmic sources (e.g. Weisskopf {\em et al.} \cite{wei02} and references 
therein; see also Ballet, this volume). 

\section{Nonthermal X-ray line production from accelerated particle
interactions}

Nonthermal electron and ion populations are expected to occur in various
astrophysical plasmas, when kinetic energy is deposited by acceleration 
processes into the tail of the distributions, at a rate that is sufficiently 
high to overcome the thermal equilibration processes. The impacts of a 
nonthermal electron tail on the ionization balance and X-ray line emission of 
such ``hybrid'' plasmas were first studied for solar flares and more recently
for supernova remnants and clusters of galaxies (e.g. Porquet {\em et al.} 
\cite{por01a} and references therein). In the following, I consider
nonthermal X-ray line production in a different context, namely from
nonthermal particles escaping from the acceleration region and interacting
in a distinct target medium. It is the most usual situation in 
laboratory experiments, but paradoxically, it has received little attention
in astrophysics. 

\subsection{Accelerated electron interactions}

Fast electrons interacting in an ambient medium can produce X-ray
lines by collisional excitation and inner-shell ionization. Nonthermal X-ray
line production by suprathermal electron interactions has been proposed to
account for the observation in impulsive solar flares of enhanced 
Fe K$\alpha$ line emission at $\sim$6.4 keV that is coincident with the hard 
X-ray burst (Zarro {\em et al.} \cite{zar92}). Recently, Valinia {\em et al.} 
(\cite{val00}) suggested that the interaction of low-energy cosmic-ray 
electrons with interstellar matter could make a significant contribution 
to the Fe K line complex observed in the spectrum of the Galactic ridge X-ray 
background. 

Here, I assume for simplicity that the interaction region is composed of
neutral atoms, such that X-ray line emission can result from inner-shell
ionization but not from collisional excitation. However, the fast electrons
can also produce continuum hard X-rays via nonthermal bremsstrahlung, which 
in turn can produce fluorescent line emission via photoionization. Thus, 
each line can generally result from a combination of collisional ionization
and photoionization, with relative contributions depending on the accelerated 
electron spectrum and on ambient medium parameters. I now discuss some 
fundamental differences between the two atomic processes. 

\begin{figure}
\includegraphics[width=5.9cm]{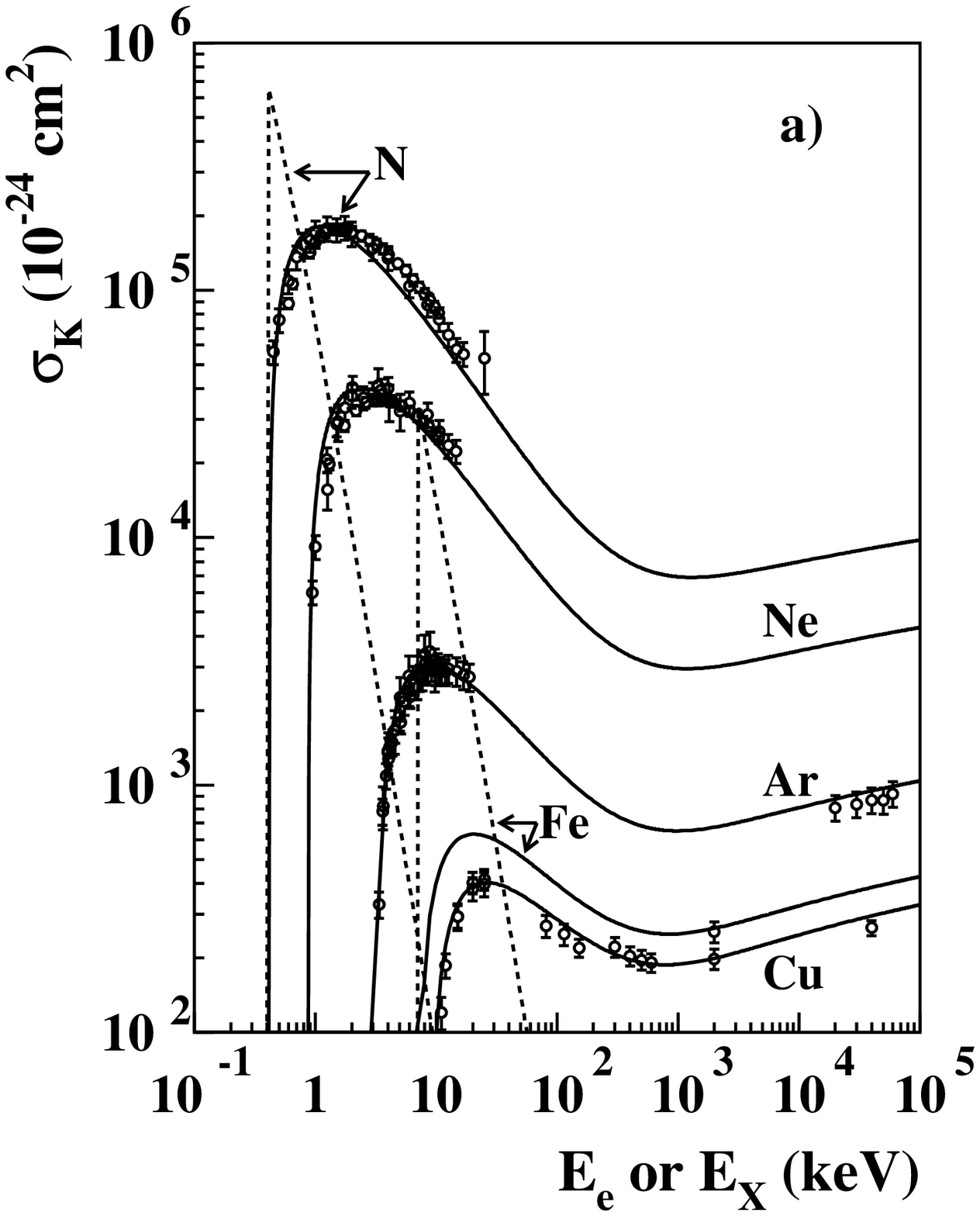}
\qquad
\includegraphics[width=5.9cm]{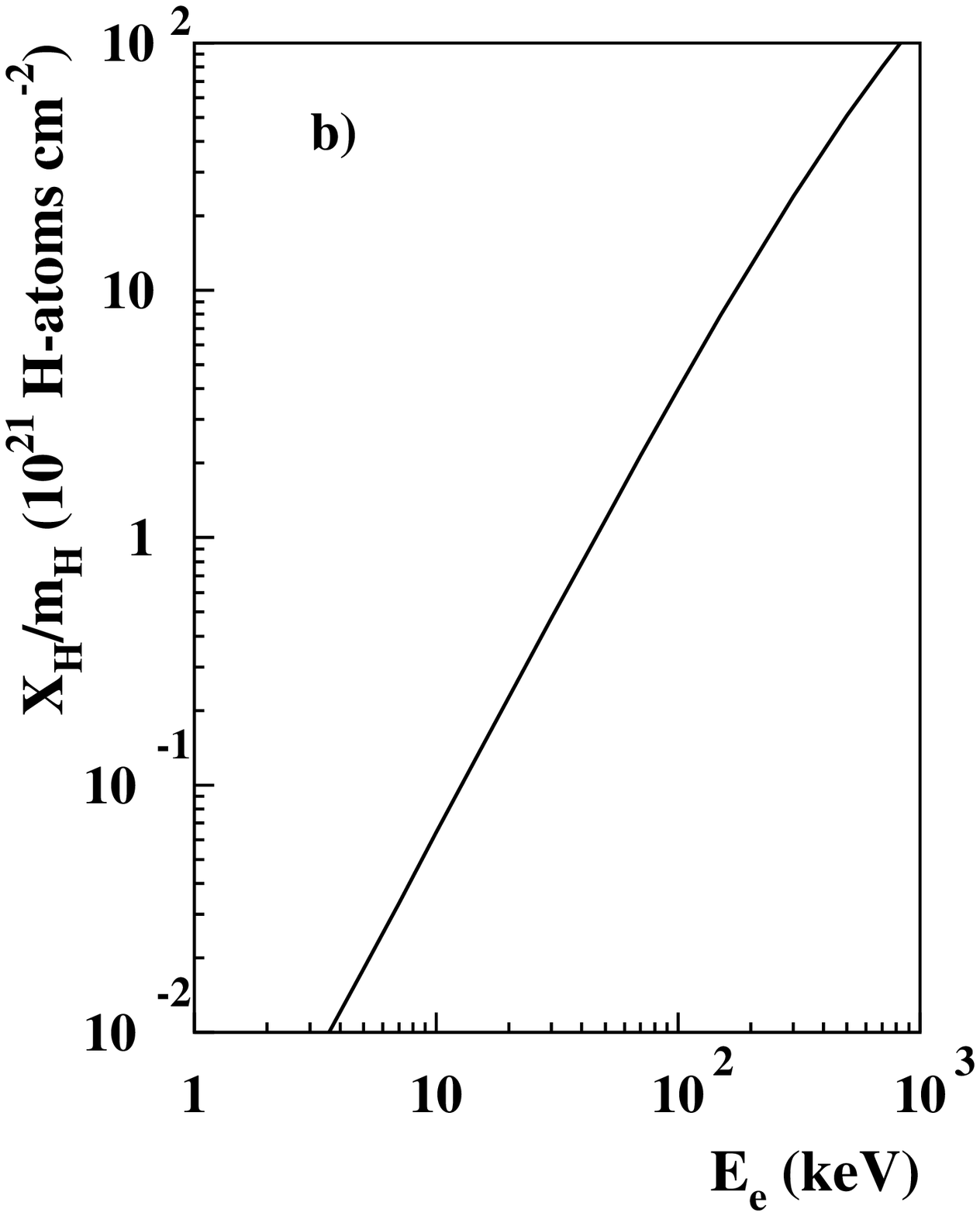}
\caption{({\it a}) Solid curves and data points: cross sections for K-shell
ionization of various atoms by electron impact. Cross sections calculated
from the semi-empirical formula of Quarles (\cite{qua76}) are compared with
data compiled in Long {\em et al.} (\cite{lon90}). Dashed curves:
cross sections for K-shell photoionization of N and Fe (Verner {\em et al.} 
\cite{ver93}). ({\it b}) Stopping range of low-energy electrons in hydrogen. 
The plotted quantity is $X_H$/$m_H$, where $X_H$ is the stopping range in 
c.g.s. units of g~cm$^{-2 }$ from Berger \& Seltzer (\cite{ber82}) and 
$m_H$ is the mass of the H-atom.}
\end{figure}

A compilation of cross sections for K-shell ionization is shown in Figure~6a.
We see that whereas the photoionization cross sections are steep functions of
energy (they vary approximately as $E_X^{-3}$ from ionization thresholds) the
cross sections for collisional ionization have a much harder energy 
dependence, in particular, a rise in the MeV region caused by 
relativistic effects (e.g. Hoffmann {\em et al.} \cite{hof79}). Thus, while 
fluorescence is essentially produced by photons with energy contained in a 
range of a few keV above the ionization threshold, ultra-relativistic 
electrons can still produce significant X-ray line emission. We also see that 
the relative contributions of collisional ionization and photoionization 
will generally depend on the target element $Z$. Numerically, the K-shell 
ionization cross section scales approximately as $z^{-4.3}$ for the 
electrons, whereas it scales about as $z^{-2.3}$ for the photons.

Propagation makes an other important difference between electron- and
photon-induced ionization. Figure~6b shows the stopping range of
low-energy electrons in a neutral H gas. Taking into account ambient He with
$(n_{He}/n_H)$=0.1 reduces the plotted range, measured in H-atoms cm$^{-2}$,
by a factor of $\sim$1.18. Thus, suprathermal electrons $<$100 keV injected
in a neutral atomic gas of solar composition would stop in 4$\times$10$^{21}$
H-atoms cm$^{-2}$. As we will see below, these electrons could produce in
such a thick target an observable Fe K$\alpha$ line emission. In comparison,
the optical depth of $>$7.1 keV X-rays with respect to photoelectric
absorption is $\tau_{abs}$$<$0.01 for $N_H$=4$\times$10$^{21}$ cm$^{-2}$, 
such that Fe K-shell ionizing photons would barely interact in the target 
region. 

As a basic introduction to nonthermal processes, let us now calculate the 
X-ray emission produced by the interaction of a steady state, suprathermal 
electron source with a thick target composed of neutral atoms with solar 
abundances. For simplicity, I assume that the nonthermal electron population
does not modify the state of the ambient medium, i.e. that ionization and
heat deposition are negligible. I further assume that the typical column 
density of the interaction region is sufficiently low to safely neglect the 
fluorescent line emission due to photoionization of the ambient atoms by the
bremsstrahlung X-rays. The differential X-ray production rate 
(photons s$^{-1}$ keV$^{-1}$) can then be written as
\begin{equation}
{dQ \over dE_X}(E_X) = \int_0^{\infty} {dN_e \over dt} (E_e) \times 
\bigg[ n_H \sum_Z a_Z \int_0^{E_e} {d\sigma_Z \over dE_X}(E_X,E_e') 
{dE_e' \over (dE/dl)(E_e')}\bigg] dE_e ~,
\end{equation}
where $(dN_e/dt)$ is the differential injection rate of suprathermal
electrons into the target region, measured in electrons s$^{-1}$ keV$^{-1}$,
$(d\sigma_Z/dE_X)$ is the differential X-ray production cross section
for electron interactions with atoms $Z$ ($\equiv$$Z^{+0}$) and $(dE/dl)$ is
the electron energy loss rate per unit path length in the ambient medium, 
measured in keV cm$^{-1}$. In this equation, the term in the brackets gives 
the monochromatic X-ray production by an electron of energy $E_e$ as it slows
down to rest in the ambient medium, which has to be integrated over the energy
distribution of the nonthermal electrons. 

\begin{figure}
\center
\includegraphics[width=8.cm]{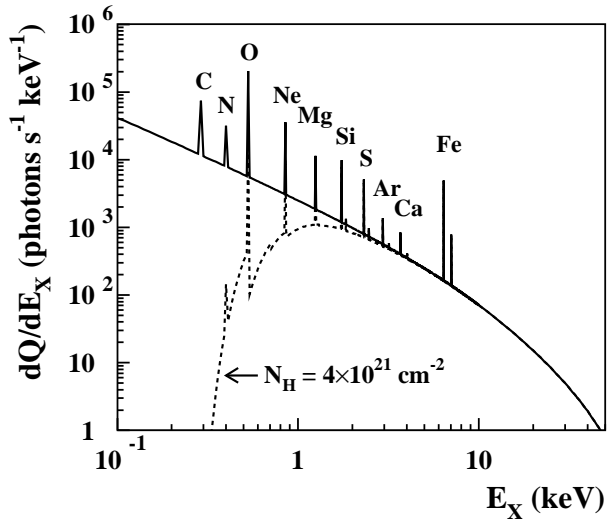}
\caption{X-ray emission produced by suprathermal electrons with the source
spectrum of eq.~(3.4) interacting in a thick target composed of a neutral 
atomic gas of solar abundance. The X-ray lines are plotted as delta functions 
in energy bins of 10 eV. The dashed line shows the effect of photoelectric 
absorption, with a H column density of 4$\times$10$^{21}$ cm$^{-2}$ 
(see text).}
\end{figure}

As H and He are by far the most abundant constituents of the 
interaction region, we have 
\begin{equation}
\bigg({dE \over dl}\bigg) \cong n_H \bigg[ m_H \bigg({dE \over dx}\bigg)_H + 
a_{He} m_{He} \bigg({dE \over dx}\bigg)_{He} \bigg]~, 
\end{equation}
where $m_H$ and $m_{He}$ are the H- and He-atom masses, and
$(dE/dx)_H$ and $(dE/dx)_{He}$ are the electron stopping powers (in
units of keV g$^{-1}$ cm$^{2}$) in ambient H and He, respectively (Berger 
\& Seltzer \cite{ber82}). Inserting eq.~(3.2) into eq.~(3.1), we see that
under the assumption of thick target interactions, the X-ray production rate
depends on the relative abundances $a_Z$ of the ambient medium constituents, 
but {\it not on the density}. 

I consider for the X-ray line production the K$\alpha$ and K$\beta$
(3$p$$\rightarrow$1$s$) lines from ambient C, N, O, Ne, Mg, Si, S, Ar, Ca and 
Fe. The corresponding cross sections can be written as
\begin{equation}
{d\sigma_Z^{Ki} \over dE_X}(E_X,E_e) = \delta(E_X-E_{Ki}) \sigma_Z^I(E_e) 
\omega_Z^{Ki}~,
\end{equation}
where $E_{Ki}$ is the energy of line K$i$ (K$\alpha$ or K$\beta$),
$\delta(E_X-E_{Ki})$ is Dirac's delta function, $\sigma_Z^I(E_e)$ is the
cross section for the K-shell ionization of atom $Z$ by an electron of energy
$E_e$ (Quarles \cite{qua76}) and $\omega_Z^{Ki}$ is the K$i$ fluorescence
yield for atom $Z$ (Firestone \cite{fir96}, appendix~F, and references 
therein). Note that $\omega_Z^{K\beta}$=0 for $z$$\leq$12 (i.e. Mg), since 
these atoms do not have 3$p$ electrons in their ground level. For the X-ray 
continuum emission, I take into account electron bremsstrahlung in ambient H 
and He only and calculate the corresponding differential cross sections from 
equation (3BN) in Koch \& Motz (\cite{koc59}), with the Elwert correction 
factor which is appropriate at nonrelativistic energies. 

The result is shown in Figure~7 (solid line) for the differential 
injection rate of suprathermal electrons 
\begin{eqnarray}{\nonumber}
{dN_e \over dt} (E_e) = 2.71\cdot10^8 \times 
E_e^{-2}~~{\rm electrons~s}^{-1}~{\rm keV}^{-1},~{\rm for~}10<E_e<100~{\rm
keV;}
\end{eqnarray}
\begin{equation}
{dN_e \over dt} (E_e) = 0,~{\rm for~}E_e<10~{\rm keV~or~}E_e>100~{\rm keV~.} 
\end{equation}

However, the calculated emission is obviously not realistic, since the
assumption that the target region is sufficiently thick to stop $<$100 keV
electrons implies that some photoelectric absorption must occur. It generally
has to be taken into account from a geometrical model of the emitting
region. In Figure~7, the dashed line spectrum is intended for providing a 
simple illustration of the effect of photoelectric absorption. It is obtained 
by multiplying $(dQ/dE_X)$ by $\exp(-N_H\sigma_{abs}(E_X))$, where
$\sigma_{abs}(E_X)$ is the absorption cross section per H-atom of a medium
of cosmic composition (Morrison \& McCammon \cite{mor83}) and $N_H$ is the
mean H column density from the site of X-ray production to the edge of the
emitting region. I adopted for this quantity the stopping range of 100 keV 
electrons in the ambient medium, $N_H$=4$\times$10$^{21}$ cm$^{-2}$.  

The normalization of the source spectrum has been chosen such that the power 
continuously deposited by the electrons in the interaction region is 
\begin{equation}
\dot{W} = \int_0^{\infty} E_e {dN_e \over dt} (E_e) dE_e = 
1~{\rm erg~s}^{-1}~. 
\end{equation}
In comparison, the total luminosity of the X-ray emission is
\begin{equation}
L_X < \int_{0.1~{\rm keV}}^{100~{\rm keV}} E_X {dQ \over dE_X}(E_X) dE_X =
4.2\times10^{-5}~{\rm erg~s}^{-1}~.
\end{equation}
This low radiation yield, $R_X$=$L_X$/$\dot{W}$, illustrates that X-ray
production by supra-thermal electrons should generally not be efficient: in a
neutral target region, almost all of the electron kinetic energy is
preferentially used to ionize ambient H and He, whereas if these two species 
are already ionized, the fast electrons will lose most of their energy by 
collective long-range Coulomb interactions. 

However, the main conclusion of this exercise is that nonthermal
bremsstrahlung should often be accompanied by observable X-ray lines below 
10 keV. In Figure~7, the K$\alpha$ lines from O, Ne, Si and Fe are the four 
most intense, with calculated equivalent widths of 355, 107, 73 and 290 eV, 
respectively\footnote{~The equivalent width is the ratio of the line intensity 
to the intensity per unit energy interval of the underlying continuum at the 
line energy.}. A significant line emission from elements of relatively low 
atomic number, such as O and Ne, could allow to distinguish X-ray line 
production by suprathermal electrons from fluorescence emission. This is 
because in the case of electron impact, the strong decrease of the 
fluorescence yields with decreasing atomic number $z$ is compensated by the 
approximate $z^{-4.3}$ dependence of the K-shell ionization cross section 
(Fig.~6a). 

\subsection{Accelerated ion interactions}

Low-energy accelerated ions can produce X-rays by a variety of atomic 
processes (Tatischeff {\em et al.} \cite{tat98} and references therein; 
Dogiel {\em et al.} \cite{dog98}). A calculated nonthermal X-ray spectrum
from accelerated ion interactions is shown in Figure~8. As for the
suprathermal electrons, I have considered a steady state, thick target
interaction model, in which accelerated ions are injected at a constant rate
into a neutral ambient medium of solar composition and produce X-rays as
they slow down to energies below the thresholds of the various reactions.
The differential X-ray production rate is calculated from an equation
similar to eq.~(3.1), but containing two more summations: one over the
elemental abundances of the fast ions and the other one over their ionization
states. I assumed that the energetic particles have the same composition
as the Galactic cosmic-ray ions at their acceleration sources (see Ramaty 
{\em et al.} \cite{ram96}, table~1). Their ionic fractions are calculated
under the assumption that the steady-state nonthermal ion population is at
the ionization equilibrium in the interaction region (see eq.~2.3 and
Tatischeff {\em et al.} \cite{tat98}). I employed the same source spectrum 
for all ion species,
\begin{equation} 
{dN_i \over dt} (E_i) \propto E_i^{-1.5}e^{-E_i/E_0}~, 
\end{equation}
which could result from strong shock acceleration in the nonrelativistic 
energy domain (Ramaty {\em et al.} \cite{ram96}). Here, $E_i$ is the kinetic 
energy of the fast ions, measured in MeV nucleon$^{-1}$, and $E_0$ is a 
parameter introduced by Ellison \& Ramaty (\cite{ell85}) for solar flare 
acceleration, which is related to the shock size and the acceleration time. 
For both clarity and simplicity, photoelectric absorption in the ambient 
medium is not taken into account. 

\begin{figure}
\center
\includegraphics[width=8.cm]{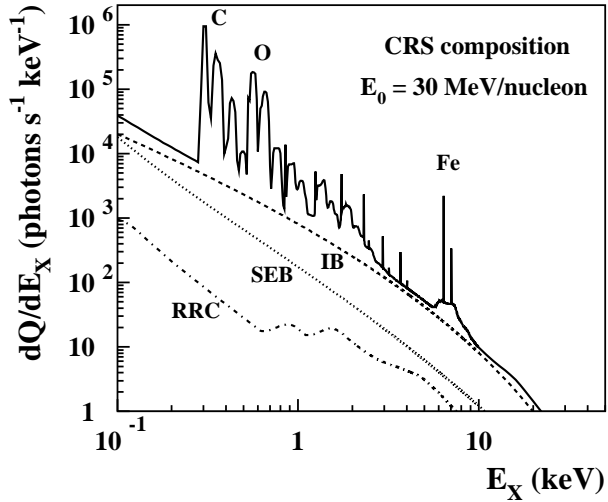}
\caption{X-ray emission produced by fast ions with cosmic-ray source (CRS)
composition and the source spectrum of eq.~(3.7) with $E_0$=30 MeV/nucleon.
The calculation is normalized to a power of 1 erg s$^{-1}$ deposited by the
accelerated particles in the interaction region. The total luminosity of the
calculated X-ray emission (see eq.~3.6) is 4.8$\times$10$^{-5}$~erg~s$^{-1}$. 
Note that photoelectric absorption in the ambient medium is not taken into 
account, such that this value is to be considered as an upper limit. IB: 
inverse bremsstrahlung; SEB: secondary electron bremsstrahlung; RRC: 
radiative recombination continuum.}
\end{figure}

Continuum X-rays are essentially produced by inverse bremsstrahlung (IB),
which is the kinematic inverse of normal bremsstrahlung, i.e. the radiation
of an electron at rest in the moving Coulomb field of a fast ion.
Thus, for nonrelativistic protons of kinetic energy $E_p$ interacting in 
stationary H, the inverse bremsstrahlung cross section is almost identical to 
the bremsstrahlung cross section for electrons of energy $(m_e/m_p)E_p$ also
interacting in stationary H\footnote{~Baring {\em et al.} (\cite{bar00}) have 
recently shown that in shocked astrophysical plasmas, IB from accelerated 
ions is generally expected to be much lower than bremsstrahlung from 
suprathermal electrons.}. Also shown in Figure~8 are the two continuum
emissions from the radiative recombinations of ambient electrons to the
K-shell of the fast ions (RRC; Dogiel {\em et al.} \cite{dog98}, Tatischeff 
\& Ramaty \cite{tat99}) and from the secondary, knock-on electrons that 
subsequently produce bremsstrahlung by interacting with the ambient atoms 
(SEB). These two emissions should generally be dominated by the IB 
production. 

The narrow lines are due to K-shell vacancy production in the ambient atoms
by the fast ions. In comparison with the similar lines produced by electron
impact or by photoionization, the lines produced by heavy ion collisions
could be shifted by several tens of electron-volts, significantly broadened 
and split up into several components, owing to multiple simultaneous 
ionizations (Garcia {\em et al.} \cite{gar73})\footnote{~These effects are
not taken into account in the present calculations. In Figure~8, the narrow 
lines are simply plotted as delta functions in energy bins of 10 eV.}. For 
example, the neutral Fe K$\alpha$ line produced by 1.9 MeV/nucleon O impacts 
is blueshifted by $\sim$50 eV in comparison with that produced by proton 
impacts, and has a full width at half-maximum (FWHM) of $\sim$100 eV (see
figure~3.55 of Garcia {\em et al.} \cite{gar73}). Such spectral signatures 
are within the capabilities of modern X-ray instruments. 

The prominent, broad line features in Figure~8 are due to atomic
de-excitations in the fast ions following electron capture by charge-transfer
reactions and collisional excitation. We have considered the K$\alpha$ and
K$\beta$ lines from H- and He-like C, N, O, Ne, Mg, Si, S and Fe (Tatischeff
{\em et al.} \cite{tat98}). These lines should generally be broad for a
steady-state nonthermal population at ionization equilibrium, because they
are produced by de-excitations of fast moving ions\footnote{~The present
calculations assume isotropic interactions, leading to the maximum possible
Doppler broadening.}. For example, the O lines in Figure~8 are produced at
$\sim$1 MeV/nucleon. The Fe broad lines are emitted at $\sim$10 MeV/nucleon
and merge together to produce a characteristic bump in the spectrum, of width
$\cong$2~keV. However, if the ambient medium is partially ionized, since 
charge exchange could then be considerably reduced, the fast ions can 
significantly slow down before capturing electrons, thus producing narrower 
lines. 

Similar line emission has recently been observed from comets,
most probably as a result of charge exchange between solar wind
particles and cometary neutral gazes (e.g. Lisse {\em et al.} \cite{lis01}).
However, the observed lines are much narrower than those of Figure~8,
because the solar wind ions remain in high charge states at low energies
(about 1 keV/nucleon), since they do not reach the ionization equilibrium in
the interplanetary space. There are as yet no astrophysical observations from 
sources outside of the solar system that would unambiguously indicate an 
origin resulting from relatively low-energy, accelerated ion interactions. 
Together with nonthermal gamma-ray line production (\S~4.2), X-ray line 
emission from charge exchange may probably be one of the most promising 
signatures of such interactions and could provide, in particular, one of the 
best ways of studying Galactic cosmic-ray ions at energies below 
100 MeV/nucleon. 

\section{Gamma-ray line production from nuclear collisions}

Let's now zoom in on the atomic nucleus, from the Angstr\"{o}m atomic scale
to the Fermi nuclear scale. Nuclear collisions can lead to gamma-ray
line emission in a variety of ways. These include the direct excitation of
nuclear levels, the production of excited secondary nuclei, the production of 
radioactive species decaying into excited states of daughter isotopes and 
the production of neutrons and positrons. The latter can originate from the
decay of both $\beta^+$-emitting radionuclei and $\pi^+$ mesons, as well as 
from the de-excitation of nuclear levels by $e^+$-$e^-$ pair emission 
(Kozlovsky {\em et al.} \cite{koz87}). Positron annihilation and the 
accompanying processes of positronium formation and annihilation are 
fundamentally important in almost all of high-energy astrophysics, as first 
pointed out by Stecker (\cite{ste69}). It is considered in \S~6. In the 
present section, I first discuss the gamma-ray line emission from 
thermonuclear reactions in high-temperature plasmas (\S~4.1), then consider 
the nonthermal gamma-ray line production from accelerated ion interactions 
(\S~4.2) and finally treat the line emission at 2.22 MeV resulting from the 
radiative capture of free neutrons by ambient protons (\S~4.3).

\subsection{Thermonuclear gamma-ray line production}

Thermonuclear reactions are responsible for the nucleosynthesis of the
elements in stars, but the associated gamma-ray production can not be
observed because stars are opaque to gamma-rays. Observable
thermonuclear gamma-ray line emission must be produced in optically thin
astrophysical plasmas with ion temperature $T_i \gsim $10$^9$ K
($kT_i \gsim $0.086 MeV). Such high-temperature plasmas are believed to exist
in the vicinity of accreting neutron stars and stellar black holes (Dubus,
this volume), as well as in central energy sources of active galactic nuclei.
In these environments, collisions of protons and $\alpha$-particles
(which are the most abundant nuclear species) with $^{12}$C and heavier ions 
can bring nuclei to excited states, which can de-excite by a
gamma-ray transition. Gamma-ray emission can also be produced through
radiative nuclear capture, for example by the reaction
$^{12}$C($p$,$\gamma$)$^{13}$N:
\begin{equation}
p + {\rm ^{12}C} \rightarrow {\rm ^{13}N} + \gamma ~.
\end{equation}

\begin{figure}
\center
\includegraphics[width=10.5cm]{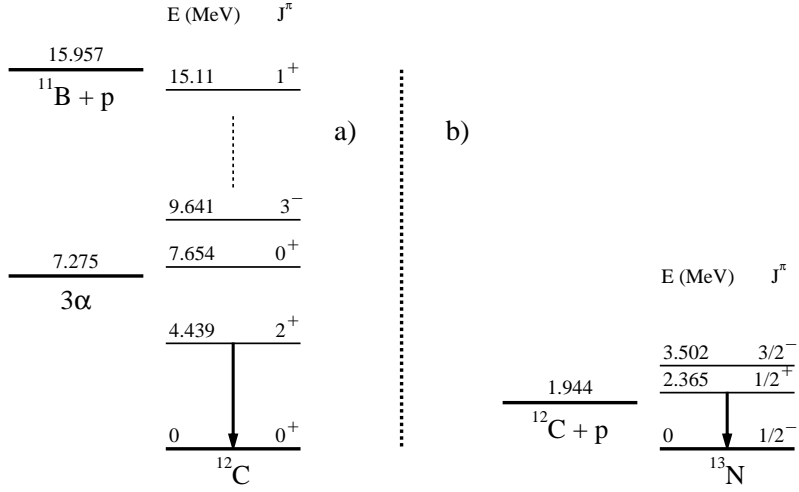}
\caption{First nuclear levels of ({\it a}) $^{12}$C and ({\it b}) $^{13}$N
($T_{1/2}$=9.965~m). Also shown are ({\it a}) the energy thresholds
for the breakup of $^{12}$C into 3 $\alpha$-particles (7.275 MeV)
and into $^{11}$B+p (15.957 MeV) and ({\it b}) the energy threshold for the
breakup of $^{13}$N into $^{12}$C+p (1.944 MeV).
The reaction $^{12}$C($p$,$\gamma$)$^{13}$N produces photons of energy (in MeV)
$E_\gamma$=1.944+$E_{cm}$, where $E_{cm}$ is the total kinetic energy of the
$^{12}$C+p interaction in its center-of-mass rest frame.} 
\end{figure}

Recently, Bloemen \& Bykov (\cite{blo97}) and Bykov {\em et al.} 
(\cite{byk99}) proposed that the collective gamma-ray line production from an 
ensemble of unresolved accreting neutrons stars could contribute to the 
diffuse Galactic emission. They were motivated by COMPTEL observations of the 
inner Galaxy, which show some evidence for an excess emission with a wide 
latitudinal distribution ($\sim$20$^\circ$ FWHM), that could correspond to 
the large-scale height of low-mass X-ray binaries (Bloemen \& Bykov 
\cite{blo97}). However, the expected emission from binary sources crucially 
depends on the accretion model. Thus, Bildsten {\em et al.} (\cite{bil92}) 
predicted very low gamma-ray line fluxes from the brightest neutron star 
X-ray binaries, if the accreting material is decelerated by Coulomb 
collisions in the neutron star atmospheres. Indeed, these authors showed that 
the infalling heavy ions would then thermalize at higher altitudes in the 
atmosphere than the accreting protons, such that they would be efficiently 
destroyed by the fast protons, until drifting to the altitude where protons 
thermalize. On the other hand, Bykov {\em et al.} (\cite{byk99}) considered 
an accretion model in which the infalling ions slow down in a 
{\it collisionless} shock produced by the outgoing radiation. In this model, 
heavy ions can coexist with protons and $\alpha$-particles in the hot 
postshock plasma for a fraction of second and thermonuclear reactions could 
then produce an observable gamma-ray line emission. Bykov {\em et al.} 
suggested that this emission could be detected with {\it INTEGRAL} from the 
nearest sub-Eddington sources. 

One of the strongest gamma-ray lines would then be at 4.438 MeV from
de-excitation of the first excited state of $^{12}$C (Fig.~9a). This level
can be populated by the inelastic scattering reactions
$^{12}$C($p$,$p$')$^{12}$C$^*_{4.439}$ and
$^{12}$C($\alpha$,$\alpha$')$^{12}$C$^*_{4.439}$, as well as by the
spallation reactions $^{16}$O($p$,$p$$\alpha$)$^{12}$C$^*_{4.439}$ and
$^{16}$O($\alpha$,2$\alpha$)$^{12}$C$^*_{4.439}$\footnote{~The difference of
1~keV between the energy of the gamma-ray line, 4.438 MeV, and the energy of
the first excited state of $^{12}$C, 4.439 MeV, is due to the recoil of the
$^{12}$C nucleus during photon emission.}. Other important lines
are those resulting from de-excitation of the most abundant heavy nuclei, e.g.
at 6.129 MeV from de-excitation of $^{16}$O. To illustrate the main
characteristics of thermonuclear gamma-ray line production, I discuss in the
following the production of the 4.438 MeV line by $p$+$^{12}$C collisions. I
also consider the radiative capture reaction (4.1), which produces a line at 
2.365 MeV at the resonance energy $E_{cm}$=0.421 MeV (Fig.~9b). 

\begin{figure}
\center
\includegraphics[width=6.3cm]{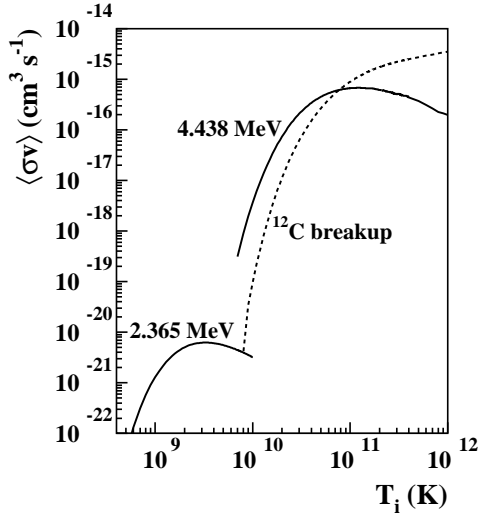}
\caption{Maxwellian-averaged reaction rates for $p$+$^{12}$C interactions, as
a function of the ion temperature.
Solid curves: productions of the 2.365 and 4.438 MeV lines from the
reactions $^{12}$C($p$,$\gamma$)$^{13}$N and
$^{12}$C($p$,$p$'$\gamma$)$^{12}$C, respectively.
Dashed curve: destruction of $^{12}$C.}
\end{figure}

I show in Figure~10 the corresponding thermonuclear reaction rates. For the
$^{12}$C($p$,$p$'$\gamma_{4.438}$)$^{12}$C reaction, I used the cross section 
tabulated by Kiener et al. (\cite{kie01}) and calculated the 
thermally-averaged rate from an equation similar to eq.~(2.4), 
\begin{equation}
<\sigma v> = \bigg({8 \over \pi \mu_{pC}} \bigg)^{1/2} 
{1 \over (kT_i)^{3/2}} \int_0^\infty E_{cm}
\sigma_{pC}(E_{cm}) e^{-E_{cm}/kT_i} dE_{cm}~,
\end{equation}
which is appropriate for an ion plasma of isothermal temperature
$kT_i \ll m_pc^2$ (e.g. Clayton \cite{cla83}). Here,
$\mu_{pC}$=$m_p m_C/(m_p$+$m_C)$ is the reduced mass of the $p$+$^{12}$C
system, $E_{cm}$ is the total kinetic energy of the reaction in its
center-of-mass rest frame and $\sigma_{pC}(E_{cm})$ is the reaction cross
section. The rate of the $^{12}$C($p$,$\gamma_{2.365}$)$^{13}$N reaction is
obtained from the analytical formula derived by Angulo et al. (\cite{ang99})
for stellar nucleosynthesis purposes, by considering only the contribution 
of the resonance at $E_{cm}$=0.421 MeV. We see that the rate of the
$^{12}$C($p$,$p$'$\gamma_{4.438}$)$^{12}$C reaction several orders of 
magnitude higher than the one of the radiative capture reaction. This is
due to the fact that inelastic scattering reactions are mostly governed by 
the strong nuclear force, whereas radiative capture reactions are essentially 
electroweak interactions. However, we also see that the 2.365 MeV line is 
produced at lower temperature, because the proton capture reaction occurs at 
lower center-of-mass energy.

The expected width of the 4.438 MeV line depends on the velocity distribution
of the excited $^{12}$C*. Following Higdon \& Lingenfelter (\cite{hig77}),
one can neglect the $p$+$^{12}$C reaction kinematics and assume that this
distribution is essentially the same as the Maxwell-Boltzmann velocity
distribution of unexcited $^{12}$C. The line FWHM in keV is then
$\sim$9$\times$($T_i$/10$^8$~K)$^{1/2}$, which gives about 90
to 300 keV for $T_i$ between 10$^{10}$ and 10$^{11}$ K. The broadening of the
2.365 MeV line is related to the Maxwellian velocity distribution of the 
center of mass of $p$+$^{12}$C interactions. One has to consider also the 
relatively large natural width of the first excited state of $^{13}$N, 
$\Gamma$=32 keV (Ajzenberg-Selove \cite{ajz91}). In a first approximation, 
the 2.365 MeV line width can then be estimated as 
FWHM$\sim$($\Gamma^2$+2.3$\times$10$^{-7}$$T_i)^{1/2}$, 
which gives about 35 to 60 keV for $T_i$ between 10$^{9}$ and 
10$^{10}$ K (Fig.~10). In addition, lines produced in the vicinity of compact 
objects can be significantly redshifted, but also broadened if the 
gravitational potential varies over the emission volume (Shvartsman 
\cite{shv72}). 

Thermonuclear destruction of the heavy species can lead to a significant
reduction in the gamma-ray line production (Guessoum \& Gould \cite{gue89a};
Guessoum \cite{gue89b}). To illustrate this point, I show in Figure~10
(dashed curve) the reaction rate for the breakup of $^{12}$C by proton
impact, which I calculated from eq.~(4.2) using for the destruction cross
section the fitting formula of Bildsten {\em et al.} (\cite{bil92}, eq.~C4).
This cross section has been estimated by subtracting the 4.438 MeV line
production cross section from the total $p$+$^{12}$C inelastic reaction cross
section, which is justified by the fact that the excited states of $^{12}$C
above the 4.439 MeV level decay primarily by particle emission\footnote{~An
exception is the 15.11 MeV level, which, because of conservation of isotopic
spin, decays essentially by $\gamma$-ray emission (Fig.~9a).}. Let $Q$ be the 
averaged number of 4.438 MeV gamma-rays emitted in a thermonuclear plasma by 
a $^{12}$C nucleus prior to destruction. We have (Bildsten {\em et al.} 
\cite{bil92})
\begin{equation}
Q \leq {<\sigma v>_\gamma \over <\sigma v>_d}~,
\end{equation}
where $<$$\sigma v$$>_\gamma$ and $<$$\sigma v$$>_d$ are the reaction rates
for gamma-ray production and $^{12}$C destruction, respectively. Thus, we see
from Figure~10 that $Q$$\ll$1 for $T_i$$\gg$10$^{11}$ K, which merely
illustrates that destruction, rather than 4.438 MeV line emission, is by
far the most probable outcome for $^{12}$C nuclei in such very 
high-temperature plasmas. Note that photon- and electron-induced reactions 
can also destroy heavy nuclei. However, Guessoum \& Gould (\cite{gue89a})
estimated that these processes should generally be negligible as 
compared with breakup by proton collisions. 

As a conclusion, a near future observation of thermonuclear gamma-ray lines
is uncertain, because the accreting heavy ions could be rapidly destroyed 
either by thermal or nonthermal processes. However, the breakup of $^4$He and 
heavier nuclei should be accompanied by a copious liberation of neutrons, that 
may eventually recombine with protons to produce a gamma-ray line emission at 
2.22 MeV (\S~4.3). This line could be produced within the high-temperature 
plasma near the compact object (Bildsten {\em et al.} \cite{bil93}) or in the 
atmosphere of the companion star (Jean \& Guessoum \cite{jea01}). In fact, 
it may be the best gamma-ray line candidate for high-energy accreting 
sources. 

\subsection{Nonthermal gamma-ray line production from accelerated ion
interactions}

The interaction of accelerated nuclei (typically $>$1 MeV/nucleon) with 
ambient matter can produce a wealth of gamma-ray lines with
energies ranging from tens of keV to about 20 MeV. This nonthermal gamma-ray
line emission is often observed from the Sun during strong solar flares
(e.g. Vestrand {\em et al.} \cite{ves99}) and it has
furnished valuable information on solar ambient abundances, density 
and temperature, as well as on accelerated particle composition, spectra and 
transport in the solar atmosphere (e.g. Ramaty \& Mandzhavidze \cite{ram00}; 
Share \& Murphy \cite{sha01}). Nonthermal gamma-ray line emission is also
expected from various astrophysical sites in which ion acceleration is 
believed to occur, such as accreting compact stellar objects, supernova
remnants and active galactic nuclei. Interactions of Galactic cosmic-ray
ions with interstellar matter should produce a diffuse emission that
might be observed in the near future. These observations would then shed 
new light on the global physical structure of the interstellar medium and on 
the sources of cosmic rays below 100 MeV/nucleon, that are not 
detected near Earth because of the solar modulation. 

The most comprehensive treatment of nuclear de-excitation gamma-ray line
emission was performed by Ramaty {\em et al.} (\cite{ram79}). The 
authors took into account more than 150 reactions leading to the production 
of an excited nucleus and evaluated their cross sections from 
laboratory measurements combined with nuclear physics theory. Figure~11 shows 
a gamma-ray line spectrum which I calculated with the corresponding 
computer code\footnote{~The code has been recently updated by taking into 
account new laboratory data allowing improved theoretical evaluations of many 
cross sections (Kozlovsky {\em et al.} \cite{koz02}). It can be downloaded 
from the 
URL: http://lheawww.gsfc.nasa.gov/users/ramaty/ViewPubs/ramaty.html~.}. I 
have considered the same steady state, thick target interaction model as for 
the nonthermal X-ray emission shown in Figure~8, in particular with the same
composition (CRS) and source spectrum (eq.~3.7) for the accelerated ions.
The luminosity of the calculated gamma-ray line emission is 
\begin{equation}
L_\gamma = \int_{0.1~{\rm MeV}}^{8~{\rm MeV}} E_\gamma {dQ \over
dE_\gamma}(E_\gamma) dE_\gamma = 1.1\times10^{-5}~{\rm erg~s}^{-1}~,
\end{equation}
which is comparable to the calculated X-ray luminosity
($L_X$$<$4.8$\times$10$^{-5}$~erg~s$^{-1}$, see Fig.~8). This should
generally be the case, except if the accelerated ion spectrum is very soft:
a high population of low-energy ions ($\sim$1~MeV/nucleon) can produce
nonthermal soft X-rays without producing a significant gamma-ray line 
emission. 

\begin{figure}
\center
\includegraphics[width=12.5cm]{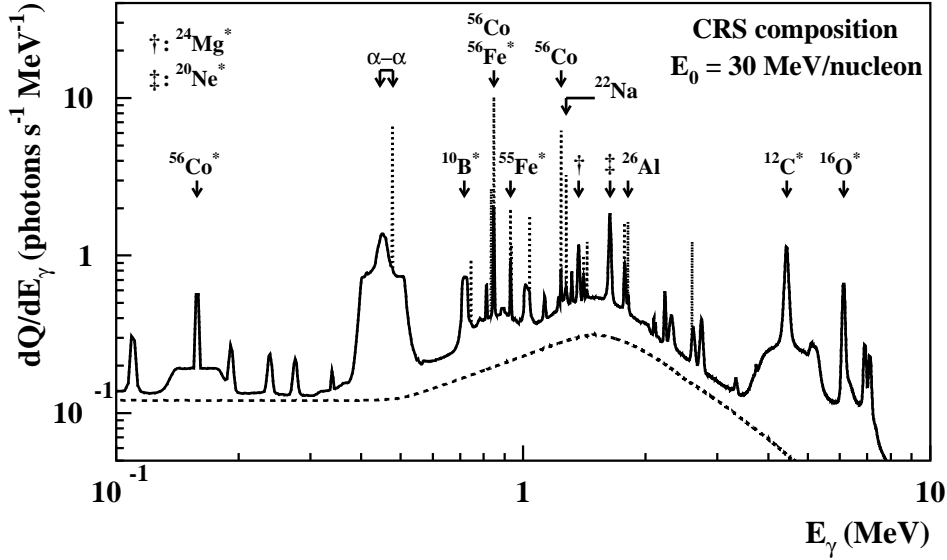}
\caption{Calculated gamma-ray line emission produced by accelerated ions with
the same composition (CRS) and source spectrum as in Figure~8, also
interacting in a neutral ambient medium of solar composition. The calculation
is normalized to the same power of 1~erg~s$^{-1}$ deposited by the accelerated
particles in the interaction region. The dashed curve shows the estimated
contribution from all unresolved gamma-ray lines in ions heavier than
$^{16}$O (see text). The very narrow dotted lines are from the decay of
long-lived radionuclei. The arrows point to the lines which are mentioned in
the text. The prompt lines are labelled with the excited nuclei from which
the gamma-rays are emitted (e.g. $^{12}$C*), whereas the delayed lines are
labelled with the parent long-lived radioisotopes (e.g. $^{26}$Al). The
so-called $\alpha$$-$$\alpha$ line feature is mainly produced by 
$\alpha$+$\alpha$ fusion reactions (see also Fig.~12).}
\end{figure}

We see in Figure~11 a combination of broad line features from
de-excitations of fast heavy ions, narrower lines from gamma-ray
transitions in heavy nuclei excited by energetic protons and
$\alpha$-particles and very narrow lines (dotted lines) from
spallation-produced long-lived radionuclei, which can come essentially to
rest in the ambient medium before decaying to an excited state of their
daughter nucleus. In addition, the prominent broad line feature at
$\sim$0.45~MeV is mainly produced by the interactions of
$\alpha$-particles with ambient $^4$He. 

All these lines are superimposed on a continuum-like emission (dashed curve),
which is due to a large number of unresolved gamma-ray lines, mostly arising 
from cascade transitions in high-lying levels of heavy nuclei. Ramaty 
{\em et al.} (\cite{ram79}) estimated the intensity and the energy 
distribution of this unresolved emission from the data of Zobel {\em et al.} 
(\cite{zob68}). These authors measured the total production of gamma-rays of 
energies $>$0.7 MeV, in interactions of protons and $\alpha$-particles with 
complex nuclei, at laboratory energies $>$10 MeV/nucleon. They found that for 
target nuclei heavier than O, the resolvable lines from de-excitation of 
the low-lying nuclear levels can account for only a fraction of the total 
gamma-ray production. In the spectrum of Figure~11, however, the unresolved 
component produces about half of the total gamma-ray emission below 3 MeV. 
New experiments could allow to specify the importance of this emission. 

The broad lines tend to overlap, such that only few well-defined features from
energetic heavy ions can be distinguished in the gamma-ray spectrum. It is
the case of the broad line feature from fast $^{12}$C*, centered at 4.4 MeV
(Fig.~11) with FWHM of $\sim$1.5 MeV, which is produced by interactions of
accelerated $^{12}$C and $^{16}$O with ambient H and He. An other example is
furnished by the broad line centered at 0.158~MeV with FWHM of $\sim$50 keV,
which is due to the de-excitation of the first excited state of $^{56}$Co,
populated by the reaction in reverse kinematics 
$^1$H($^{56}$Fe,$^{56}$Co*)$n$.  

Intense narrow lines are due to excitations of low-lying nuclear levels in
abundant ambient ions, e.g. at 0.847~MeV from $^{56}$Fe*, 1.37~MeV from
$^{24}$Mg*, 1.63~MeV from $^{20}$Ne*, 4.44~MeV from $^{12}$C* and 6.13~MeV
from $^{16}$O*. Other important narrow lines are those arising from
transitions in the spallation products of abundant ambient nuclei, e.g. at
0.718~MeV from $^{10}$B*, which is produced by the spallation of
ambient $^{12}$C and $^{16}$O, and at 0.931~MeV from $^{55}$Fe*, which is
essentially produced by the reaction $^{56}$Fe($p$,$p$$n$)$^{55}$Fe*. The narrow
lines produced in a gaseous ambient medium are generally broadened by the
recoil velocity of the excited nucleus and their FWHM is about 0.5-5\% of the
transition energy. However, some lines produced in interstellar dust grains
can be very narrow, because some of the excited nuclei can stop in solid
materials before emitting gamma-rays\footnote{~This is not taken into account 
in the present calculations, for which I assumed that the ambient medium is
devoid of dust grains.} (Lingenfelter \& Ramaty \cite{lin77}). It requires 
that (i) the mean life time of the excited nuclear level or of its radioactive
nuclear parent is longer than the slowing down time of the excited nucleus
in the grain material and (ii) the mean distance from the site of the nuclear
interaction to the grain edge is higher than the stopping range of the
excited nucleus in the grain. The most promising candidates are from the
de-excitation of the levels of $^{56}$Fe at 0.847~MeV ($T_{1/2}$=6.1~ps) and
1.238~MeV ($T_{1/2}$=640~fs), of $^{24}$Mg at 1.37~MeV ($T_{1/2}$=1.35~ps),
of $^{28}$Si at 1.78~MeV ($T_{1/2}$=475~fs) and of $^{16}$O at 6.13~MeV
($T_{1/2}$=18.4~ps). Most of the interstellar Fe, Mg and Si, which are
refractory elements, could be contained in dust grains, whereas about half
of the interstellar O could be in grains (e.g. Savage \& Sembach 
\cite{sav96}). The detection of these very narrow lines, maybe with the
{\it INTEGRAL} spectrometer (SPI), would provide unique information on the
composition, size and spatial distribution of the interstellar grains. 

Both the broad and the narrow lines could in principle have complex
profiles that generally depend on the composition, energy and
angular distributions of the accelerated particles\footnote{~In the updated 
nuclear de-excitation line code of Ramaty {\em et al.}, the accelerated 
particles are assumed to have an isotropic distribution in the interaction 
region.}, the composition of the ambient medium and the mechanism of the 
nuclear reactions producing the excited nuclei. In particular, the detailed 
calculation of line shapes requires the knowledge of the angular and energy 
distributions of the recoil nuclei and the angular distribution of the 
emitted gamma-rays. Ramaty {\em et al.} (1979) estimated the shapes of the 
various lines from a limited number of available laboratory data together 
with simplifying theoretical assumptions. More recent works have shown that 
fine spectroscopic analyses of the gamma-ray line profiles could give 
valuable information on the directionality of the accelerated particles in 
the interaction region (Kiener {\em et al.} 2001 and references therein). 

In Figure~11, the most intense narrow lines from spallation-produced
radionuclei are at 0.478~MeV from the decay of $^{7}$Be ($T_{1/2}$=53.3~d), at
0.847 and 1.238~MeV from the decay of $^{56}$Co ($T_{1/2}$=77.3~d) and at 1.275
MeV from the decay of $^{22}$Na ($T_{1/2}$=2.6~y). These lines are very narrow
only if the density of the ambient medium is high enough for the radionuclei
to come essentially to rest prior to decay. The line at 1.809~MeV from
$^{26}$Al decay should generally be very narrow, because the time to stop 
energetic ($<$100 MeV/nucleon) $^{26}$Al in a medium of $\sim$1 H-atom 
cm$^{-3}$ is shorter than its half-life ($T_{1/2}$=7.4$\times$10$^5$ y). But
$^{7}$Be ions, which are mostly produced by the reaction
$^4$He($\alpha$,$n$)$^7$Be with recoil energy $>$2 MeV/nucleon, stop in a 
medium with solar abundances in less than 53.3~days only if the H density 
exceeds 10$^5$ cm$^{-3}$. 

\begin{figure}
\center
\includegraphics[width=12.5cm]{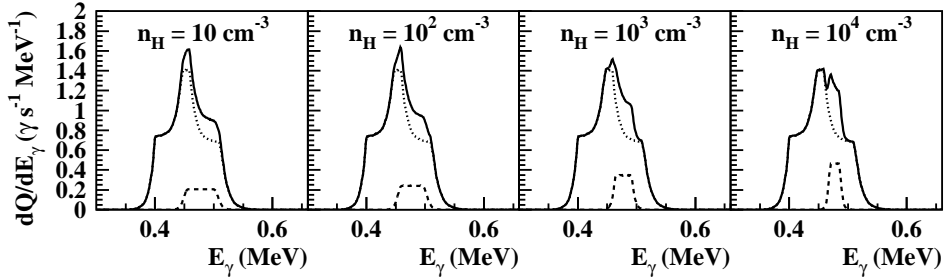}
\caption{Profiles of the gamma-ray line emission at $\sim$0.45~MeV, for 4
values of the H density in the interaction region. The total emission
(solid curves) is the sum of the prompt lines from the reactions
$^4$He($\alpha$,$n$$\gamma_{0.429}$)$^7$Be and
$^4$He($\alpha$,$p$$\gamma_{0.478}$)$^7$Li (dotted curves) and 
the delayed line at 0.478 MeV from $^7$Be decay (dashed curves).}
\end{figure}

Figure~12 shows detailed profiles of the gamma-ray emission at $\sim$0.45~MeV 
for lower values of the ambient H density  (Tatischeff {\em et al.} 
\cite{tat01}). The delayed line at 0.478~MeV from $^{7}$Be decay is 
superimposed on a prompt emission from the reactions
$^4$He($\alpha$,$n$$\gamma_{0.429}$)$^7$Be and
$^4$He($\alpha$,$p$$\gamma_{0.478}$)$^7$Li. The two prompt lines
merge (for the assumed isotropic distribution of energetic
$\alpha$-particles) thus producing this characteristic emission feature,
whose importance in gamma-ray astrophysics was first pointed out by
Kozlovsky \& Ramaty (\cite{koz74}). We see that the width of the delayed
line decreases as the H density increases, because $^{7}$Be
ions decay at lower energies in a denser propagation region. This provides a 
potential density measurement, which could allow, in particular, to 
determine if low-energy cosmic-ray ions can penetrate into the core of  
interstellar molecular clouds. 

In summary, interactions of energetic particles with ambient material 
produce a great variety of nuclear de-excitation lines, that could provide a 
unique tool to study the acceleration processes at work in the Galaxy and 
beyond, and also to examine the ultimate nature of the interaction regions. 
In particular, observations of diffuse Galactic emission would permit the
first measurements of the isotopic composition of the interstellar medium, 
independent of the ionization and molecular states of its constituents.  

\subsection{The radiative capture of neutrons}

\begin{figure}
\center
\includegraphics[width=8.5cm]{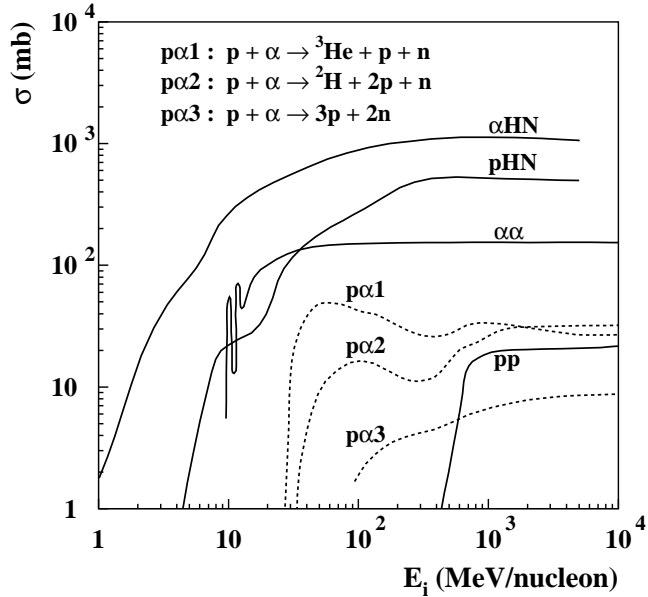}
\caption{Neutron production cross sections in units of
millibarn$\equiv$10$^{-27}$~cm$^{-2}$, as a function of the proton or
$\alpha$-particle bombarding energy. Here, $pp$, $p\alpha$,
$\alpha \alpha$, $pHN$ and $\alpha HN$ indicate neutron production in
proton--hydrogen, proton--helium, alpha particle--helium, proton--heavy
nuclei and alpha particle--heavy nuclei interactions, respectively. The
cross sections of the three neutron-producing reactions in
proton--helium interactions are shown separately (dashed curves). The
$pHN$ and $\alpha HN$ reaction cross sections take into account all isotopes 
equal to or heavier than $^{12}$C (see text). All the cross sections 
include the neutron multiplicity.}
\end{figure}

Free neutrons can be produced by the breakup of all isotopes equal to or
heavier than $^2$H, either by thermonuclear or nonthermal reactions. Their
subsequent radiative capture by ambient nuclei can produce many specific
gamma-ray lines, among which the line at 2.22 MeV from the reaction
$^1$H($n$,$\gamma$)$^2$H is expected to be the strongest in most
astrophysical sites. 

I show in Figure~13 the cross section for important neutron-producing
reactions. The $pHN$ and $\alpha HN$ reaction cross sections are from Hua \&
Lingenfelter (\cite{hua87a}; see also Ramaty {\em et al.} \cite{ram75}). 
They are defined as  
\begin{eqnarray}
\sigma_{pHN}={1 \over \sum_k a_k} \sum_i a_i \sigma_{pi} \\
\sigma_{\alpha HN}={1 \over \sum_k a_k} \sum_i a_i \sigma_{\alpha i}
\end{eqnarray}
where $a_i$ is the solar abundance of isotope $i$ with respect to $^1$H,
$\sigma_{pi}$ (resp. $\sigma_{\alpha i}$) is the cross section for neutron
production in collisions of protons (resp. $\alpha$-particles) with isotope
$i$ and the summations are over all isotopes equal to or heavier than
$^{12}$C ($\sum_k a_k$=1.6$\times$10$^{-3}$). The cross sections of the
$pp$, $p\alpha$ and $\alpha \alpha$ reactions are from Murphy {\em et al.}
(\cite{mur87}; see also Meyer \cite{mey72}). These reactions should
generally be the dominant neutron-producing processes in interactions of 
accelerated particles of energy $>$10 MeV/nucleon and in thermonuclear 
plasmas. However, spallation of heavy nuclei plays an important role in
solar flares, because the accelerated ions have generally a rapidly 
decreasing energy distribution. 

\begin{figure}
\center
\includegraphics[width=7.cm]{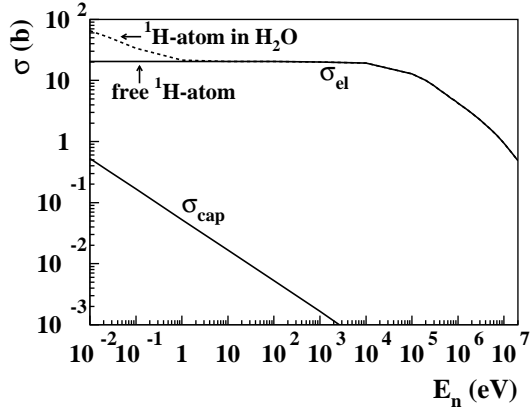}
\caption{Cross sections for the elastic scattering of a neutron on a free
$^1$H-atom and on a $^1$H-atom contained in a water molecule
(dashed curve), and for the radiative capture reaction
$^1$H($n$,$\gamma$)$^2$H, as a function of the neutron energy
(from Horsley \cite{hor66}). The radiative capture cross section satisfies 
$\sigma_{cap}(E_n)$=2.44$\times$10$^{-6}$/$\beta_n$ (in b) for $E_n$$<$1~keV, 
where $\beta_n$ is the neutron velocity relative to that of light.}
\end{figure}

Shown in Figure~14 are cross sections for $n$+H interactions. We see that the
neutrons have a much larger probability to scatter on ambient
$^1$H-atoms than to be captured. They lose on average about
half of their kinetic energy in each elastic scattering. Thus, most of them
get thermalized before being possibly captured, if they haven't decayed
($n$$\rightarrow$$p$+$e^-$+$\bar{\nu}_e$) or escaped from the interaction 
region. In a medium of H density $n_H$, the mean time for radiative 
capture is given by 
\begin{eqnarray}
\tau_{cap}&=&{1 \over n_H \sigma_{cap} \beta_n c}  \\
&=&{1.4\times10^{19}~{\rm s} \over n_H~{\rm (cm^{-3})}} ~{\rm for~} E_n < 1~{\rm keV~,}
\end{eqnarray}
as $\sigma_{cap}(E_n)$=2.44$\times$10$^{-30}$/$\beta_n$ (in cm$^2$) for 
$E_n$$<$1~keV (Fig.~14). This characteristic time must be compared with the 
mean time for neutron decay:
\begin{equation}
\tau_{dec} = {T_{1/2} \over \ln(2)} = 886~{\rm s,}  
\end{equation}
where $T_{1/2}$=613.92$\pm$0.55~s is the neutron half-life (Groom {\em et al.} 
\cite{gro00}). Thus, significant gamma-ray line emission at 2.22~MeV can only
be produced in an ambient medium of H density exceeding
$\sim$10$^{16}$ cm$^{-3}$, i.e. in a {\it stellar environment}. The line 
should generally be narrow if the only broadening mechanism is thermal. If 
it is the case, its FWHM exceeds the energy resolution of a Germanium 
detector ($\sim$2.5~keV at 2.2~MeV) only if the ambient medium temperature 
is$\gsim$5$\times$10$^6$~K. 

As pointed out by Wang \& Ramaty (\cite{wan74}), ambient $^3$He can
constitute an important non-radiative sink for the neutrons. Indeed, the
cross section for the reaction $^3$He($n$,$p$)$^3$H has the same form at low
energies as that of the $^1$H($n$,$\gamma$)$^2$H reaction
($\sigma(E_n)$~$\propto$~$1/\beta_n$), but is
$\sim$1.6$\times$10$^4$ times larger, which is to be compared with the 
$^1$H/$^3$He solar abundance ratio, 7.23$\times$10$^4$ (Anders \& Grevesse 
\cite{and89}). In fact, this coincidence has allowed an original 
determination of the $^3$He photospheric abundance, from observations of the 
2.22~MeV line produced in solar flares (Hua \& Lingenfelter \cite{hua87b}). 

The COMPTEL all-sky survey at 2.22~MeV revealed one point-like 
feature (GRO J0332-87) at the Galactic coordinates
($\ell$,$b$)=(300.5$^\circ$,-29.6$^\circ$) (McConnell {\em et al.}
\cite{mcc97}; Sch\"onfelder {\em et al.} \cite{sch00}). The most interesting
known source in the COMPTEL location error box is RE J0317-853, which is one 
of the hottest ($\approx$50,000$^\circ$~K) and most highly magnetized
($\approx$340~MG) white dwarfs, residing at a distance of only 
$\sim$35~pc. Observations with {\it INTEGRAL} are expected to provide a new
insight into the origin of this yet mysterious emission.  

\section{Gamma- and X-ray lines from stellar nucleosynthesis}

Observations of gamma-ray lines produced by the decay of radioisotopes
synthesized in stars have long been recognized as a powerful tool for
studying the nucleosynthesis processes at work in stellar interiors and
explosions (Clayton \cite{cla82} and references therein; Kn\"odlseder, this
volume). However, among the hundreds of potential radioactive emitters of
gamma-ray lines, only a few of them are suitable for astronomical
observations. First, nucleosynthesis sites are generally optically thick to
gamma-rays, such that the appropriate radioisotopes must have a sufficiently
long half-life (typically $T_{1/2}$$>$1~day) so as to decay only after having 
left the dense production environment. Additionally, their production yields 
must be high enough for the decay activities to be observable by gamma-ray 
instruments, which, in the present state of the gamma-ray line astronomy, 
practically excludes all the radioisotopes beyond the Fe peak. 

\begin{table}
\caption{Star-produced radioisotopes relevant to gamma-ray line astronomy.}
\vspace{2mm}
\begin{tabular}{|c|c|c|c|c|}
\hline
{\bf Iso-} & {\bf Production} & {\bf Decay} & {\bf Half-} &
{\bf $\gamma$-ray energy (keV)} \\
{\bf tope} & {\bf sites$^{\rm a}$} & {\bf chain$^{\rm b}$} &
{\bf life$^{\rm c}$} & {\bf and intensity$^{\rm d}$} \\\hline
$^7$Be & Nova & $^7$Be~$\stackrel{\epsilon}{\longrightarrow}$~$^7$Li*
& 53.3~d & 478~(0.11) \\\hline
$^{56}$Ni & SNIa, CCSN &
$^{56}$Ni~$\stackrel{\epsilon}{\longrightarrow}$~$^{56}$Co*
& 6.075~d & 158~(0.99), 812~(0.86) \\
& & $^{56}$Co~$\stackrel{\epsilon(0.81)}{\longrightarrow}$~$^{56}$Fe*
& 77.2~d & \underline{847}$^{\rm e}$~(1), \underline{1238}~(0.67) \\\hline
$^{57}$Ni & SNIa, CCSN &
$^{57}$Ni~$\stackrel{\epsilon(0.56)}{\longrightarrow}$~$^{57}$Co*
& 1.48~d & 1378~(0.82) \\
& & $^{57}$Co~$\stackrel{\epsilon}{\longrightarrow}$~$^{57}$Fe*
& 272~d & \underline{122}~(0.86), \underline{136}~(0.11) \\\hline
$^{22}$Na & Nova &
$^{22}$Na~$\stackrel{\beta^+(0.90)}{\longrightarrow}$~$^{22}$Ne*
& 2.61~y & \underline{1275}~(1) \\\hline
$^{44}$Ti & SNIa, CCSN &
$^{44}$Ti~$\stackrel{\epsilon}{\longrightarrow}$~$^{44}$Sc*
& 60.0~y & \underline{68}~(0.93), \underline{78}~(0.96) \\
& & $^{44}$Sc~$\stackrel{\beta^+(0.94)}{\longrightarrow}$~$^{44}$Ca*
& 3.97~h & \underline{1157}~(1) \\\hline
$^{26}$Al & CCSN, WR &
$^{26}$Al~$\stackrel{\beta^+(0.82)}{\longrightarrow}$~$^{26}$Mg*
& 7.4$\cdot$10$^5$~y & \underline{1809}~(1) \\
 & AGB, Nova & & & \\\hline
$^{60}$Fe & CCSN &
$^{60}$Fe~$\stackrel{\beta^-}{\longrightarrow}$~$^{60}$Co*
& 1.5$\cdot$10$^6$~y & 59~(0.02$^{\rm f}$) \\
& & $^{60}$Co~$\stackrel{\beta^-}{\longrightarrow}$~$^{60}$Ni*
& 5.27~y & 1173~(1), 1332~(1) \\\hline
\end{tabular}
{\vspace{1mm} \small

{\hspace{0.4cm} $^{\rm a}$ Sites which are believed to produce observable
$\gamma$-ray line emission. Nova: classical nova; SNIa: thermonuclear
supernova (type Ia); CCSN: core-collapse supernova; WR: Wolf-Rayet star;
AGB: asymptotic giant branch star.}

{\hspace{0.4cm} $^{\rm b}$ $\epsilon$: orbital electron capture. When an
isotope decays by a combination of $\epsilon$ and $\beta^+$ emission, only 
the most probable decay mode is given, with the corresponding fraction in 
parenthesis.}

{\hspace{0.4cm} $^{\rm c}$ Half-lifes of the isotopes decaying by 
$\epsilon$ are for the neutral atoms (see text).}

{\hspace{0.4cm} $^{\rm d}$ Number of photons emitted in the $\gamma$-ray
line per radioactive decay.}

{\hspace{0.4cm} $^{\rm e}$ The underlined $\gamma$-ray lines are those for
which a positive detection has been reported at the time of writing.}

{\hspace{0.4cm} $^{\rm f}$ It is noteworthy that the first excited
(isomeric) state of $^{60}$Co at 59~keV decays with 98\% probability by the 
emission of a conversion electron.}
}
\end{table}

Table~1 lists the most important cosmic radioactivities for gamma-ray line
observations. The astrophysical sites in which these isotopes are synthesized
are discussed in the review paper of Diehl \& Timmes (\cite{die98}).
The nuclear data are from Firestone (\cite{fir96}, and references therein for
the $\epsilon$-$\beta^+$ decay branchings), except for the nuclei of mass
number A=44, 56 and 57, for which
I used the more recent data evaluations of Cameron \& Singh (\cite{cam99}),
Junde (\cite{jun99}) and Bhat (\cite{bha98}), respectively. The half-life of
the radioisotopes listed in the Table should be compared with the
characteristic time scale between two nucleosynthesis events which produced
them in the Galaxy, $\sim$1-2 weeks for novae, $\sim$40~yr for CCSN and
$\sim$300~yr for SNIa (e.g. Prantzos \cite{pra99} and references therein).
Thus, we see that we expect the decays of the long-lived $^{26}$Al and
$^{60}$Fe nuclei to produce diffuse gamma-ray line emissions, resulting from
the superposition of numerous Galactic sources. On the other hand, the
gamma-ray line activities of $^{7}$Be, $^{56}$Ni, $^{57}$Ni and $^{44}$Ti
should be observed in individual transient events. The case of $^{22}$Na is 
in a way intermediary: the line emission at 1.275~MeV could be observed from
nearby individual ONe novae ($<$1.1~kpc with SPI, Hernanz et al.
\cite{her01}), but also from a broad region towards the Galactic center,
resulting from the cumulated production of few tens of active sources. 
Evidence for this latest unresolved emission has recently been reported by 
the COMPTEL team (Iyudin {\em et al.} \cite{iyu02}). 

In Table~1, six of the radioisotopes decay mainly or exclusively by orbital
electron capture ($\epsilon$). As pointed out by Mochizuki {\em et al.}
(\cite{moc99}) for $^{44}$Ti, one has to be cautious when deducing
the production yields of these isotopes from the delayed observations of their
gamma-ray activities, since their decay rates depend on their ionization
states. Thus, in the interstellar medium, fully ionized
$\epsilon$-radioisotopes can only decay by highly improbable nuclear electron
capture from continuum states, such that they are almost stable as compared
with the age of the universe\footnote{~This is not the case in stellar
interiors because of the much higher density of continuum electrons. For
example, the half-life of $^7$Be against capture of free electrons in the
core of the Sun is $\approx$90~d (Bahcall \& Moeller \cite{bah69}), which is
by coincidence comparable to the measured value for neutral atoms.}.
Mochizuki {\em et al.} (\cite{moc99}) suggested that the ionization of
$^{44}$Ti behind the reverse shock of the Cas A supernova remnant (SNR) could 
have significantly reduced its decay rate, thus leading to a lower mass of
ejected $^{44}$Ti than previously inferred from the detected 1.157~MeV line 
flux. However, a more realistic model for the ejecta shows that the effect 
may not exceed a few percent (Laming \cite{lam02}). 

The decay of an $\epsilon$-radioisotope can also produce X-ray line emission,
when the electron vacancy in the daughter atom (primarily in its K-shell) is
filled by a radiative transition\footnote{~The adjustment of the nucleus is
generally much faster than the atomic rearrangement, such that the atomic
transition usually occurs in the daughter element.}. Leising (\cite{lei01})
has recently examined the exciting possibility of detecting cosmic 
radioactivities in X-rays. Many of them are not gamma-ray line emitters, 
because they decay directly to the ground state of their daughter nucleus. The 
author listed 17 potentially important $\epsilon$-radioisotopes, among which 
the most promising for detection are $^{55}$Fe ($T_{1/2}$=2.73~yr), $^{44}$Ti
($T_{1/2}$=60~yr), $^{59}$Ni ($T_{1/2}$=7.6$\times$10$^4$~yr) and $^{53}$Mn
($T_{1/2}$=3.74$\times$10$^6$~yr). The $\epsilon$ decay of $^{55}$Fe could be
observed through the emission of the Mn K$\alpha$ line, from very young SNRs 
such as SN1987A. The
X-ray line emission which accompanies the $^{44}$Ti decay should be hunted 
for in the two SNRs from which gamma-ray line emission at 1.157~MeV has 
already been observed, Cas A and RX J0852-4622. The detection of this X-ray
line emission would constitute an interesting complement to the gamma-ray
line measurements, because it would allow to determine the present ionization
state of Ti in the two SNRs. Finally, the decays of $^{59}$Ni and $^{53}$Mn, 
producing K$\alpha$ lines at 6.92 and 5.41~keV, respectively, could be 
observed from very close and reasonably young (i.e. not too extended) SNRs 
and also as diffuse Galactic emissions\footnote{~A 
diffuse emission in the Co K$\alpha$ line at 6.92~keV is also 
expected from the decay chain
$^{60}$Fe~$\stackrel{\beta^-}{\longrightarrow}$~$^{60}$Co$^*_{59~{\rm keV}}$
$\stackrel{IT}{\longrightarrow}$~$^{60}$Co ($IT$: internal transition), since
the de-excitation of $^{60}$Co$^*_{59~{\rm keV}}$ is essentially due to the 
conversion of a K-shell electron (see Table~1). The expected Galactic flux is
\begin{equation}
F_X = F_\gamma {\alpha(K) \over 1 + \alpha} \omega_{Co}^{K\alpha} \cong 
1.8 \times 10^{-5}~{\rm photons}~{\rm s}^{-1}~{\rm cm}^{-2}~,
\end{equation}
where $F_\gamma$$\cong$6.5$\times$10$^{-5}$ $\gamma$ s$^{-1}$ cm$^{-2}$
is the predicted flux for the 1.173~MeV line from $^{60}$Co decay (Timmes 
\& Woosley \cite{tim97}), $\alpha$=48 and $\alpha(K)$=40 are the total and
K-shell conversion coefficients for the 59~keV transition (King \cite{kin93})
and $\omega_{Co}^{K\alpha}$=0.33 is the K$\alpha$ fluorescence yield for 
Co. In comparison, Leising (\cite{lei01}) predicted that the
decay of $^{59}$Ni should produce a 6.92~keV line flux from the central 
steradian of the Galaxy of $\sim$1.7$\times$10$^{-4}$ photons s$^{-1}$
cm$^{-2}$.}. As Leising pointed out, it could be difficult to distinguish the
electron capture line emission from astrophysical sources which are generally 
expected to be very luminous in X-rays. In particular, detailed models of 
SNRs are required to determine if the expected lines could be unambiguously
identified among the other thermal and nonthermal X-ray line emissions. 
However, one can reasonably hope that in a few cases, the clear detection of 
X-ray lines from cosmic radioactivities will constitute an important 
complement to the gamma-ray line studies of stellar nucleosynthesis. 

\section{Positron annihilation radiation}

Positron-electron annihilation radiation can be produced in a variety of 
astrophysical sites. The diffuse Galactic emission at 511 keV is the 
brightest gamma-ray line of cosmic origin. It is produced by the steady-state 
annihilation of $\sim$1.4-4.3$\times$10$^{43}$ positrons s$^{-1}$ (depending 
on the spatial distribution of the emission, Kinzer {\em et al.} 
\cite{kin01}). A large fraction of these positrons could result from the 
$\beta^+$ decay of $^{56}$Co and $^{44}$Sc (synthesized as $^{56}$Ni and 
$^{44}$Ti in supernovae, see Table~1), with a smaller contribution from the 
$\beta^+$ decay of $^{26}$Al (Chan \& Lingenfelter \cite{cha93}; Dermer \& 
Murphy \cite{der01}). Milne {\em et al.} (\cite{mil02}) have recently 
estimated that these three isotopes could supply 30-50\% of the Galactic 
positrons. The decay of $\pi^+$ and radioisotopes produced by 
cosmic-ray interactions in the interstellar medium could make an additional 
contribution of a few percent (e.g. Ramaty \& Lingenfelter \cite{raml79}). 

Positrons are also produced in classical novae, from the decay of
radioisotopes synthesized during the thermonuclear runaway. The main
$\beta^+$-radionuclei synthesized in novae are $^{13}$N ($T_{1/2}$=9.965~min),
$^{18}$F ($T_{1/2}$=109.8~min), $^{22}$Na ($T_{1/2}$=2.61~h) and $^{26}$Al
($T_{1/2}$=7.4$\times$10$^5$~y). The decay of the latter two isotopes could
inject a small but non-negligible amount of positrons in the interstellar
medium, but $^{13}$N and $^{18}$F have too short half-lifes, so that most of 
the positrons should annihilate in the relatively dense material ejected by 
the nova explosion. The resulting gamma-ray emission could be detected with 
SPI from relatively nearby novae during the first day after the
explosion\footnote{~The detectability of this emission crucially depends on
the uncertain rates of the reactions $^{18}$F($p$,$\gamma$)$^{19}$Ne and
$^{18}$F($p$,$\alpha$)$^{15}$O, which are responsible for the thermonuclear 
destruction of $^{18}$F (Coc {\em et al.} \cite{coc00}).} 
(Hernanz et al. \cite{her01}). 

Positron production can also occur in accreting compact objects, from 
the pair production reaction $\gamma + \gamma ~{\rightarrow}~ e^- + e^+$, 
from the materialization of photons in strong magnetic fields
$\gamma + \overrightarrow{B} ~{\rightarrow}~ e^- + e^+ + \overrightarrow{B}$ 
and from other less important processes such as 
$\gamma + X ~{\rightarrow}~ e^- + e^+ + X$, where $X \equiv e^-$ or nucleus. 
A significant fraction of these positrons could annihilate near the compact 
objects, thus producing a localized and possibly time-variable emission. The 
annihilation radiation from relativistic plasmas is discussed by Marcowith 
(this volume). Here, I consider the emission produced by the injection of 
nonthermal positrons in various astrophysical environments. 

\begin{figure}
\center
\includegraphics[width=11.cm]{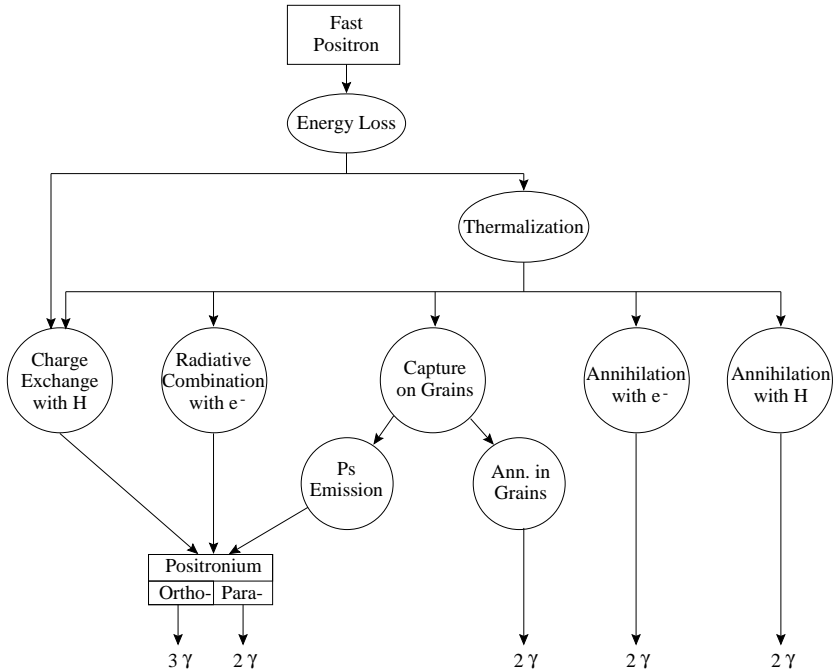}
\caption{Main processes leading to gamma-ray emission from positron 
annihilation (adapted from Guessoum {\em et al.} \cite{gue91}, fig.~1).}
\end{figure}

Important processes leading to gamma-ray emission from positron annihilation
are shown in Figure~15 (see Guessoum {\em et al.} 
\cite{gue97} for discussion of the various processes and update of the cross 
sections). Positrons produced by the decay of radioactive nuclei
have initial kinetic energies typically ranging from several hundred keV to
few MeV. They slow down mainly by Coulomb interactions with free electrons
and by excitations and ionizations of neutral atoms, and their characteristic
slowing-down time is $\sim$10$^5$/$n$~years, where $n$ is the ambient medium
density in units of cm$^{-3}$ (Guessoum {\em et al.} \cite{gue91}).
Positrons produced by other processes generally have higher initial energies.
Having lost the bulk of its kinetic energy, a positron can either form a
positronium (Ps) atom in flight by charge exchange with an atom or a molecule 
(at $\lsim$50~eV in neutral H or H$_2$, Bussard {\em et al.} \cite{bus79}) or
become thermalized with the free electrons. Thermal positrons either
annihilate directly with free or bound electrons or form Ps atoms through
radiative recombination and charge exchange (Fig.~15). As first pointed out 
by Zurek (\cite{zur85}), interstellar dust could play a significant role in
annihilation of thermal positrons despite its low abundance, since
its projected surface area ($\sim$10$^{-21}$~cm$^2$ per Galactic H-atom) can
be larger than the cross sections for the other processes leading to positron
annihilation. Collisions of thermal positrons with dust grains can lead to 
the reflection of the positrons from the grains, to the annihilation of the
positrons within the grains and to the formation and subsequent escape of 
Ps atoms back to the gas. The contribution of these processes to 
gamma-ray emission crucially depends on the grain properties, including
their compositions, sizes, charges and abundances in the ambient medium. 

The positronium atom, which is the simplest hydrogen-like system, is
preferentially formed in the state of principal quantum number $n$=1, either
as a triplet ($^3S_1$) in which the spins of the electron and positron are
parallel, or as a singlet ($^1S_0$) in which they are
antiparallel\footnote{~The formation of a Ps atom in an excited state can
produce a Ly$\alpha$ line at 2430~\AA~and other lines in the UV, IR and radio
bands, that could provide an other way to observe positron annihilation from
astrophysical sources (e.g. Wallyn {\em et al.} \cite{wal96}).}. These
states are called, respectively, orthopositronium and parapositronium, by
analogy with the spectral designations for hydrogen. Parapositronium
annihilates with a lifetime $\tau_p$=1.25$\times$10$^{-10}$~s into
two photons of 511~keV in its rest frame, but orthopositronium decays with
$\tau_o$=1.4$\times$10$^{-7}$~s into three photons, which form a
characteristic continuum at energies below 511 keV (Ore \& Powell
\cite{ore49}). Ps atoms can be quenched before annihilation if the ambient
medium density is $\gsim$10$^{13}$~cm$^3$. 

Let $I_{e^+}$ be the total number of positrons which annihilate in a given 
source, and $I_{2\gamma}$ and $I_{3\gamma}$ the number of photons emitted 
from this source in the 511 keV line and in the ortho-Ps continuum, 
respectively. Since the probability for formation of the lepton atom in the 
triplet state is three times the probability for its formation in the singlet 
state, we have 
\begin{equation}
I_{2\gamma}=I_{e^+} \times [ 2\times (1/4) \times f_{Ps} + 2\times
(1-f_{Ps})]
\end{equation}
\begin{equation}
{\rm and~~}I_{3\gamma}=I_{e^+} \times 3 \times (3/4) \times f_{Ps}~,
\end{equation}
where $f_{Ps}$ is the fraction of positrons that annihilate via Ps
formation. Combining these two equations gives $f_{Ps}$ in terms of the 
observable quantity ($I_{2\gamma} / I_{3\gamma}$):
\begin{equation}
f_{Ps}={2 \over 2.25(I_{2\gamma} / I_{3\gamma})+1.5~}.
\end{equation}

\begin{figure}
\center
\includegraphics[width=10.cm]{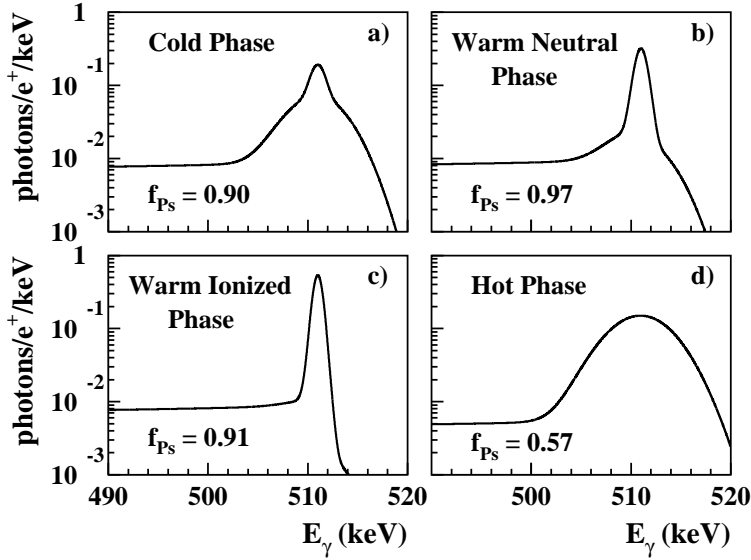}
\caption{Calculated gamma-ray spectra of the positron annihilation radiation
produced in the various phases of the interstellar medium. Processes
involving interstellar dust are not taken into account. The calculations are
normalized to one positron annihilation in each phase. $f_{Ps}$: fraction of 
positrons that annihilate through Ps formation.}
\end{figure}

To illustrate how an observation of the 511~keV line profile and the
determination of $f_{Ps}$ can provide unique information on the annihilation
environment, I show in Figure~16 calculated spectra of the radiation
produced by positron annihilation in the four phases of the
interstellar medium. I used equation~(9) of Guessoum {\em et al.}
(\cite{gue91}) with the rates of the various processes and the corresponding
line widths given in that paper, except for the widths of the lines produced
by radiative recombination and direct annihilation of thermal positrons, for
which I used the more accurate results of Crannell {\em et al.}
(\cite{cra76}; see Guessoum {\em et al.} \cite{gue97}). For simplicity, I
neglected the processes involving interstellar dust. 

We see in panel ({\it a}) that in the cold phase, the line is
relatively broad and the fraction of positrons which annihilate via Ps
formation is high. Indeed, in this neutral environment, most of the
nonthermal positrons form Ps in flight by charge exchange and the immediate
annihilation of para-Ps produces a broad line (6.4~keV FWHM, Brown
{\em et al.} \cite{bro84}). On this component is superimposed a
narrower line due to the direct annihilation of 10\% of the positrons with
bound electrons. In the warm neutral phase (of temperature $T$=8000~K and
ionization fraction $\zeta$=$n_{e^-}$/($n_{e^-}$+$n_H$)=0.15, McKee \&
Ostriker \cite{mck77}), the fraction of positrons forming Ps in flight is
reduced to 22\%. Most of the positrons get thermalized and form Ps by charge
exchange just above the energy threshold (6.8~eV in the center of mass for
$e^+$+H${\rightarrow}$Ps+$p$). The annihilation line has a
FWHM of 1.5~keV (Bussard {\em et al.} \cite{bus79}). In the warm ionized
phase ($T$=8000~K, $\zeta$=0.68), the 511~keV line is even narrower
(1.1~keV FWHM), because a significant fraction of thermal positrons
form Ps through radiative recombination, which does not have an energy
threshold. Finally, in the hot phase of the interstellar medium
($T$=4.5$\times$10$^5$~K), all the positrons get thermalized and the
annihilation line should be broad ($\approx$7.4~keV FWHM) due to
the thermal motion of the the center of mass of the $e^+$-$e^-$ pairs 
which annihilate. Crannell {\em et al.} (\cite{cra76}) estimated
the width of the 511~keV line from both para-Ps annihilation following
radiative recombination and direct annihilation with free electrons as
FWHM=11$\times$($T$/10$^6$~K)$^{1/2}$~keV. However, the line width could be
significantly reduced if annihilation on interstellar dust is important 
(Guessoum {\em et al.} \cite{gue91}). The results shown in Figure~16 are in 
good agreement with the recent calculations of Dermer \& Murphy 
(\cite{der01}), except for the cold phase, for which they found the narrow 
line component from positron annihilation with bound electrons to be 
negligible.

Harris {\em et al.} (\cite{har98}) have measured, with the TGRS instrument
(Germanium detector) on the {\it WIND} spacecraft, the width of the 511 keV 
line from the Galactic center region to be narrow, 
FWHM=1.81$\pm$0.54$\pm$0.14~keV, and the Ps fraction to be large, 
$f_{Ps}$=0.94$\pm$0.04, in good agreement with the most recent OSSE 
result: $f_{Ps}$=0.93$\pm$0.04 (Kinzer {\em et al.} \cite{kin01}). The narrow 
line width suggests that most of the positrons can not penetrate deeply into 
molecular clouds and that annihilation in the hot interstellar phase is not 
predominant, unless dust grains sprinkled into the hot plasma are important 
sites of annihilation. However, the Ps fraction would then be very low 
($<$0.5, Guessoum {\em et al.} \cite{gue91}), in conflict with the measured 
values. Thus, these observations favor a scenario in which the Galactic 
positrons annihilate predominantly in the warm, partially ionized 
interstellar medium. The spectroscopic-imaging capability of {\it INTEGRAL} 
should allow a better determination of the nature and the spatial 
distribution of the emitting regions. This would shed new light on the 
sources of the positrons and on the thermodynamic properties of the 
annihilation media. 

\vspace{0.5cm}
{\noindent \small I would like to thank Jean Ballet, J\"urgen Kiener, 
David Lunney and Alexandre Marcowith for critical reading of the manuscript.}


\end{document}